%% file: main.tex
\documentclass[fleqn,usenatbib]{mnras}             
\usepackage{newtxtext,newtxmath}                   
\usepackage[T1]{fontenc}
\usepackage{lipsum}
\usepackage{hyperref}
\usepackage{bm}
\usepackage[utf8]{inputenc}

\DeclareRobustCommand{\VAN}[3]{#2}
\let\VANthebibliography\thebibliography
\def\thebibliography{\DeclareRobustCommand{\VAN}[3]{##3}\VANthebibliography}

\usepackage{graphicx}	                           
\usepackage{amsmath}	                           
\usepackage{caption}
\usepackage{csquotes}                              
\usepackage{hyperref}
\usepackage{nicefrac,xfrac}
\usepackage[dvipsnames]{xcolor}
\usepackage{tikz}
\usepackage[normalem]{ulem}
\newcommand{\swatch}[1]{\tikz[baseline=-0.6ex] \node[fill=#1,shape=rectangle,draw=black,thick,minimum width=5mm,rounded corners=2pt](){};}

\newcommand{\orcid}[1]{\href{https://orcid.org/#1}{\includegraphics[scale=0.05]{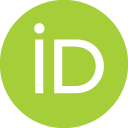}}}

\title[SMBHs in simulated jellyfish galaxies]{Jellyfish galaxies
with the IllustrisTNG simulations -- 
Supermassive black hole activity in dense environments with ram-pressure stripped satellites}

\author[S. Kurinchi-Vendhan et al.]
       {Shalini Kurinchi-Vendhan,$^{\orcid{0009-0007-8974-9016} 1,2}$
        \thanks{E-mail: kurinchi@mpia.de}
        Eric Rohr,$^{\orcid{0000-0002-9183-5593} 1,3,4}$
        Annalisa Pillepich,$^{\orcid{0000-0003-1065-9274} 1}$
        Elad Zinger,$^{\orcid{0000-0002-6316-3996} 5,1}$
        \newauthor
        Mohammadreza Ayromlou,$^{\orcid{0000-0003-3783-2321} 6}$ and 
        Gandhali D. Joshi $^{\orcid{0000-0003-4665-0765} 7}$
\\ \\
$^{1}$ Max-Planck-Institut f\"ur Astronomie, K\"onigstuhl 17, Heidelberg D-69117, Baden-W\"urttemberg, Germany
\\
$^{2}$ California Institute of Technology, 1200 E. California Boulevard, Pasadena, CA 91125, USA 
\\
$^{3}$ Universit\"{a}t Heidelberg, Exzellenz-Cluster STRUCTURES, Philosophenweg 12, 69120 Heidelberg, Germany
\\
$^{4}$ Max-Planck-Instiut f\"ur Astrophysik, Karl-Schwarzschild-Strasse 1, 85748 Garching bei M\"unchen, Germany
\\
$^{5}$ Centre for Astrophysics and Planetary Science, Racah Institute of Physics, The Hebrew University, Jerusalem 91904, Israel
\\
$^{6}$ Argelander-Institut f\"ur Astronomie, Auf dem H\"ugel 71, D-53121 Bonn, Germany
\\
$^{7}$ Department of Physics, Institute for Computational Cosmology, Durham University, South Road, Durham DH1 3LE, UK
}

\date{Accepted 2025 July 23. Received 2025 July 19; in original form 2025 June 05}
\pubyear{2025}

\begin{document}
\label{firstpage}
\pagerange{\pageref{firstpage}--\pageref{lastpage}}
\maketitle

\begin{abstract}
\input{sections/sec0_abstract}
\end{abstract}

\begin{keywords}
galaxies: formation -- galaxies: evolution -- galaxies: active -- galaxies: clusters -- methods: numerical
\end{keywords}


\input{sections/sec1_introduction}
\input{sections/sec2_methods}

\input{sections/sec3_results}

\input{sections/sec4_discussion}
\input{sections/sec5_conclusion}


\section*{Acknowledgements}                                         %
SK is grateful to the U.S. Fulbright Scholarship for funding her work on this project.  SK is also thankful to the authors of \citet{peluso_exploring_2022} and the GASP team for providing the stellar mass data of observed jellyfish galaxies. This work is also supported by the DFG under Germany’s Excellence Strategy EXC 2181/1-390900948 (the Heidelberg STRUCTURES Excellence Cluster). MA is supported at the Argelander Institute f\"ur Astronomie through the Argelander Fellowship. The flagship simulations of the IllustrisTNG project used in this work have been run on the HazelHen Cray XC40-system at the High Performance Computing Center Stuttgart as part of project GCS-ILLU of the Gauss centres for Supercomputing (GCS). The TNG50 simulation was realized with compute time granted by the GCS under the GCS Large-Scale Project GCS-DWAR
(2016; PIs Nelson/Pillepich). Additional analysis was conducted using the the Vera cluster of the Max Planck Institute for Astronomy (MPIA), which is operated by the Max Planck Computational Data Facility (MPCDF). This publication uses data generated via the Zooniverse platform, the development of which is funded by generous support, including a Global Impact Award from Google, and by a grant from the Alfred P. Sloan Foundation. We wish to extend our thanks to the team at Zooniverse and the thousands of volunteers who invested their time and effort to assist us in this project.

\section*{Data Availability and Software}                           %

Data directly related to the figures of this publication are available on request from SK. Data from the IllustrisTNG simulations are publicly available \url{www.tng-project.org} \citep{nelson_first_2019}. The LBE data, including LBE density, velocity, and ram pressure, will be shared upon reasonable request to MA. Information about the ``Cosmological Jellyfish'' Zooniverse data can be found through \citet{zinger_jellyfish_2023} and the Zooniverse website (\url{https://www.zooniverse.org/projects/apillepich/cosmological-jellyfish}). The data was analysed and visualized with \textsc{Python}, utilizing the \textsc{NumPy} \citep{harris2020array}, \textsc{SciPy} \citep{2020SciPy-NMeth}, and \textsc{Matplotlib} \citep{Hunter:2007} packages. Finally, this work makes extensive use of the NASA Astrophysics Data System and the arXiv.org e-Print archive.

\bibliographystyle{mnras}                            
\typeout{}
\bibliography{references/refs}

\appendix                                                           %
\input{sections/sec6_appendix}


\bsp	          
\label{lastpage}
\end{document}

%% file: sections/sec0_abstract.tex
\noindent Jellyfish galaxies are extreme examples of how galaxies can transform due to dense environmental effects. These satellite galaxies suffer from ram-pressure stripping, leading to the formation of their distinctive gaseous tails. Some recent observational studies find that jellyfish galaxies are more likely to host active galactic nuclei (AGN) compared to central galaxies of the same mass, suggesting a link between ram pressure and supermassive black hole (SMBH) accretion. We use the IllustrisTNG cosmological-magnetohydrodynamical simulations, namely TNG50 and TNG100, to explore the presence of AGN in jellyfish galaxies with $M_{\rm{stellar}}\simeq10^{9.5-10.8}\,\rm{M}_\odot$ at redshift $z=0$ from the Zooniverse ``Cosmological Jellyfish'' citizen-science project. Compared to central galaxies, jellyfish are more likely to host an AGN ($L_{\rm AGN}\geq10^{44}\,\mathrm{erg\,s^{-1}}$) particularly at high stellar masses ($M_{\rm stellar}\gtrsim10^{10}\,\mathrm{M_\odot}$). Jellyfish are also more likely to host an AGN than satellites of the same mass, largely because many satellite galaxies are gas-poor and therefore have lower SMBH accretion rates. Compared to non-jellyfish satellites with similar gas content, jellyfish typically undergo stronger ram pressure and have higher central gas densities along with lower central gas sound speeds, although these effects are smaller at lower stellar masses ($M_{\rm stellar}\lesssim10^{10}\,\mathrm{M_\odot}$). Together with case studies of individual galaxies, our population analysis indicates that ram pressure can play a key role in fuelling AGN activity in a large fraction of jellyfish, where gas compression can lead to intense episodes of AGN feedback and star formation. Thus, it is essential to consider both environmental and secular processes for a more complete picture of satellite galaxy evolution.

%% file: sections/sec1_introduction.tex

\section{Introduction} \label{SEC:Introduction}

Exploring the formation and evolution of galaxies in a cosmological context is key to understanding how the Universe came to be, with its diverse range of galaxy sizes, colors, and morphologies. Galaxy clusters are a hotbed for this type of research: as galaxies interact with each other and the surrounding gas, the fundamental properties of their structure and content can change dramatically. An extreme of these transformations are galaxies that exhibit bright tails of gas trailing behind their main stellar bodies. Named for their uncanny resemblance to jellyfish (in observations \citealp{Ebeling_2014} and \citealp{Fumagalli_2014}; and more recently in simulations \citealp{yun_jellyfish_2019}), these galaxies are undergoing hydrodynamic stripping of their gas as they move rapidly through dense gaseous environments.

The main types of processes behind galaxy evolution are external influences from the environment and internal star formation-driven processes and feedback. Through near encounters with surrounding galaxies in a dense environment, such as in a galaxy group or cluster, gravitational interactions can alter all components of the galaxy. Galaxy strangulation \citep{Peng_2015} prevents the flow of gas from the intracluster medium into the galaxy as it enters the larger gravitational potential of a cluster for the first time; tidal stripping \citep{Gnedin_2003} pulls away gas and stellar material; and galaxy harassment \citep{Moore_1996} radically changes the distribution of matter in the galaxies during interactions with the central galaxy or other satellites. These environmental processes can lead to asymmetries, tidal tails, and other warped morphologies in satellites. In the case of ``jellyfish'' galaxies, ram-pressure stripping \citep{Gunn_Gott_1972, Balogh_2000} is the source of their extraordinary appearance. It occurs when a satellite travels through a dense cosmic medium and experiences the force exerted by the surrounding gas. If the ram pressure surpasses the gravitational restoring force of the galaxy, it removes gas from the outer layers and forms tails.

Hundreds of jellyfish have now been observed by looking for the effects of environmental stripping on galaxies. They are found through the visual identification of extended tails or ``tentacles'' of gas emanating away from the disk of an otherwise unperturbed stellar body. These studies encompass a broad range of wavelengths including radio \citep{Miley_1972, Merluzzi_2024}, neutral hydrogen \citep{Kenney_2004}, ionized H$\alpha$ gas \citep{Gavazzi_2001, Cortese_2006,Boselli_2016, Gavazzi_2018}, molecular gas \citep{Jachym_2014, Jachym_2017, Verdugo_2015}, the ultraviolet regime \citep{Smith_2010, George_2024}, and X-rays \citep{Machacek_2006, Sun_2006}. Observations of jellyfish tails span from single galaxies within clusters like Virgo or Coma in the aforementioned studies, to large sample surveys such as the LOFAR Two-Meter Sky Survey \citep[LoTSS;][]{Roberts_2021} and the OSIRIS Mapping of Emission-Line Galaxies \citep[OMEGA;][]{Roman-Oliveira_2019}, which each detect tens of jellyfish. Perhaps the most recent ambitious search for jellyfish galaxies is the GAs Stripping Phenomena in Galaxies with MUSE program \citep[GASP;][]{poggianti_gasp_2017}, with a statistical sample of $\sim100$ jellyfish galaxies. 

So far, most observational studies have focused on detecting signatures from the environment in the jellyfish properties. However, internal processes that happen inside a galaxy itself are crucial to its evolution as well, such as feedback from stellar winds and supernovae, and the central supermassive black hole. SMBHs are found to reside in nearly all massive galaxies ($M_{\rm{stellar}} \geq 10^{10} \: \rm{M}_\odot$) in the local Universe, suggesting a strong connection to host galaxy properties as they co-evolve through cosmic time \citep{Kormendy_2013, McConnell_Ma_2013}. On smaller scales, stellar feedback can lead to heating shocks and outflows that redistribute the galactic gas reservoir \citep{Ciotti_1991}. In the energetic phase of SMBHs, active galactic nuclei or AGN feedback is meanwhile thought to be fundamental to regulating the amount of gas and star formation in galaxies through releasing intense levels of radiation, driving powerful winds, and leading to the ionization and expulsion of gas via relativistic jets and outflows \citep{Bower_2006, Croton_2006, Fabian_2012}. Theoretical studies have demonstrated that AGN feedback is key to solving important questions in galaxy evolution, including how they stop actively forming stars \citep[e.g.][]{Di_Matteo_2023}.

While the impact of SMBH feedback in massive, central galaxies is known to be important, the relative significance of SMBHs in satellite galaxies stills needs a consensus. \cite{Dressler_1985} first proposed that the frequency of AGN is higher in the field compared to cluster galaxies based on emission-line diagnostics. But additional emission-line studies show that AGN prefer low-density environments \citep{Kauffmann_2004}; other radio observations also suggest that AGN can be more prevalent in high-density environments depending on the properties of the host cluster \citep{Best_2012, Sabater_2013, Lopes_2017, Man_2019}; and along with some X-ray studies conclude that the incidence of AGN does not depend on the environment at all \citep{Miller_2003, Martini_2007,  Sivakoff_2008, Arnold_2009, Linden_2010, Amiri_2019}. The outcomes of these observational studies could differ due to their various methods for characterizing galaxy environments and AGN. Consequently, the presence of SMBHs is less understood when dense environments enter the scene.

Galaxies are complicated systems and, as a result, internal and external factors are often at interplay and can occur simultaneously. It has been speculated that ram pressure from a gaseous environment can compress the interstellar medium of a satellite galaxy, leading to enhanced star formation \citep[][in observations]{Vulcani_2018, Vulcani_2024, Roberts_2022, Roberts_2023} and causing gas to flow inward toward the galactic center and be accreted by the SMBH. While theoretical studies have mostly focused on the implications of star formation alone \citep{Schulz_2001, Tonnesen_2009, Tonnesen_2012, Ramos_Martinez_2018}, ram pressure may also drive AGN in this scenario. At the same time, feedback from star formation and SMBH activity can in turn lower the binding energy of the central gas reservoir and cause outflows, potentially facilitating the formation of jellyfish tails \citep[][again from the perspective of stellar feedback]{Garling_2022}. The energy from AGN feedback injected into the interstellar medium can thus strengthen the efficiency of ram pressure, significantly contributing to the distinctive jellyfish-like structures.

Jellyfish galaxies can provide insight into both internal feedback and environmental mechanisms as they are not only sites of strong ram-pressure stripping, but also can be hosts of SMBHs according to recent observational literature. \cite{poggianti_ram-pressure_2017} find that five out of seven jellyfish from a GASP sample of highly-stripped galaxies host an AGN based on the Baldwin–Phillips–Terlevich \citep[BPT;][]{Baldwin_1981} diagram with hints of ionization, outflows, and star formation quenching from feedback \citep{Radovich_2019, George_2019}. Extending to a larger statistical sample of more than a hundred galaxies, \cite{peluso_exploring_2022} detects AGN in almost a third of jellyfish---half, considering only the most massive objects. Their sample selects for star-forming galaxies, and thus it remains an open question whether the high AGN fraction that \cite{peluso_exploring_2022} observes actually has a relation to ram-pressure stripping itself.

\begin{figure*}
    \centering
    \vspace{0.5em}
    \includegraphics[width=0.95\linewidth]{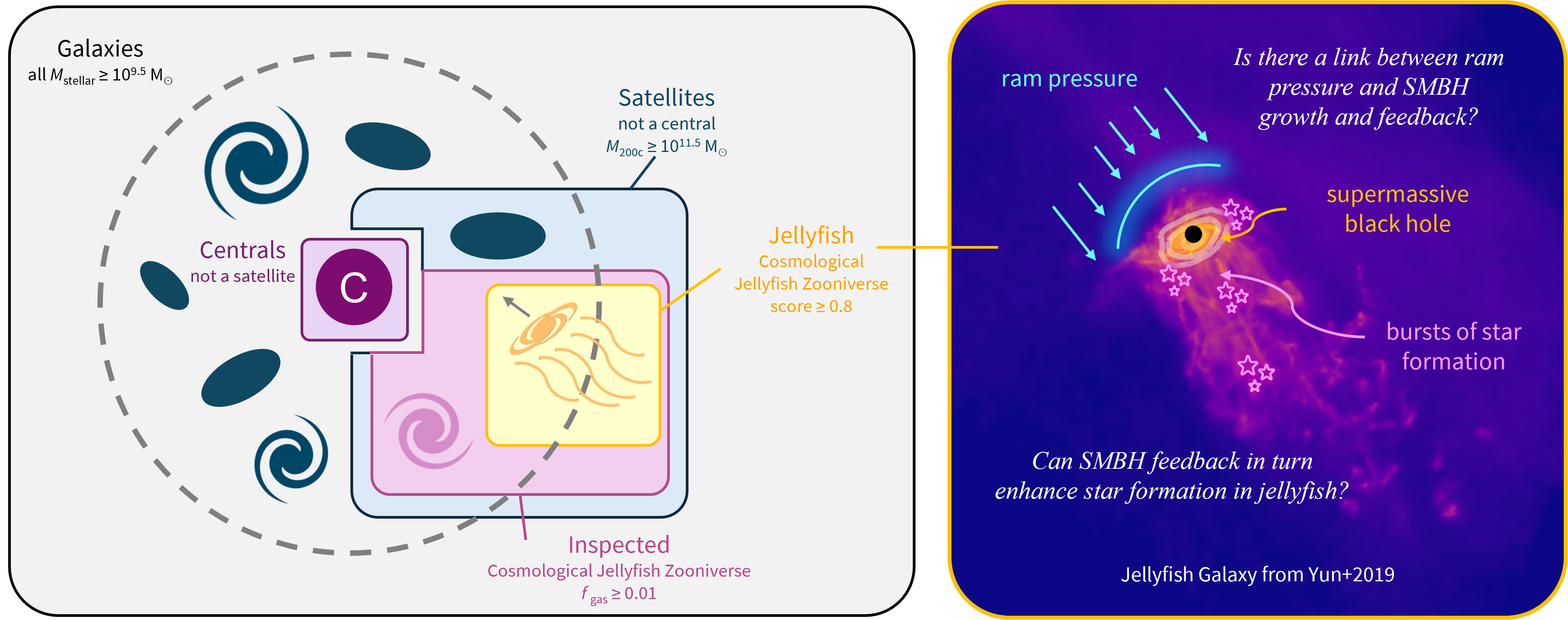}
    \vspace{0.5em}
    \caption{\textit{\textbf{Infographic of IllustrisTNG galaxies and cosmological jellyfish.}} In a selection of all galaxies from TNG50 and TNG100 with $M_\mathrm{stellar} \geq 10^{9.5} \: \mathrm{M}_\odot$ (gray), we consider satellites (blue) in hosts of total mass $M_\mathrm{200c} \geq 10^{11.5} \: \mathrm{M}_\odot$ that may experience environmental effects compared to centrals (purple). The ``Cosmological Jellyfish'' Zooniverse project \protect\citep{zinger_jellyfish_2023} inspected a subset of gaseous satellites (pink; $f_{\rm gas} = M_{\rm gas, \: total} / M_{\rm stellar, \: 2 \: \text{half-mass}} \geq 0.01$) to identify jellyfish. Jellyfish galaxies (yellow) are named for the gaseous tails that emanate from their galactic disk due to experiencing ram-pressure in a dense cluster environment. The poster-image on the right shows the gas density profile of a TNG100 from ~\protect\citet{yun_jellyfish_2019}. The influence of the environment and SMBH feedback can possibly interact in ram-pressure stripped satellites, with effects on the jellyfish galaxy properties.}
    \label{FIG:Sample_Selection}
\end{figure*}

While the GASP collaboration finds strong evidence for a high incidence of nuclear activity among ram-pressure stripped jellyfish galaxies, the OMEGA survey \citep{Roman-Oliveira_2019} reports that only five out of seventy jellyfish are hosts of AGN according to H$\alpha-$EW \textit{vs.} [\ion{N}{ii}]$/$H$\alpha$ diagrams \citep{Cid_Fernandes_2011}, and thus they are not able to draw a link between the two phenomena. Using \textit{Chandra} observations, \citet{Tiwari_2024} do not find evidence for enhanced AGN activity in a sample of almost thirty ram-pressure stripped galaxies from four clusters, though they note that X-ray measurements of SMBHs are known to suffer from gas and dust obscuration. \citet{Koulouridis_2024} find that ram pressure can sometimes trigger AGN depending on their distance from the cluster centre. As this brief summary of recent results suggests, the true AGN fraction in jellyfish galaxies is still open to debate from the perspective of observations, considering their various sample sizes and criteria for detecting AGN activity. Whether SMBHs are more common in satellites compared to galaxies in the field is important to exploring if the environment might drive nuclear activity.

To understand the tension across different observational studies, it is useful to turn to theoretical methods. Whereas observations can provide a picture of galaxy evolution at one point in in cosmic time, simulations allow us to trace the histories of galaxies over billions of years. \cite{Akerman_2023} perform hydrodynamic wind-tunnel simulations of a jellyfish with an SMBH and propose that ram pressure increases the flow of gas to the galaxy center, supporting an enhanced frequency of AGN in stripped galaxies. However, wind-tunnel simulations represent the case of a single galaxy. Moreover, the detailed physics of black hole accretion under ram-pressure stripping is not well understood. More samples of galaxies must be studied and controlled galaxy samples as a point of reference must be carefully designed, necessitating larger theoretical experiments. 

Modern large-volume cosmological simulations of galaxies have largely succeeded in modelling the evolution of structure and star formation in galaxies over cosmic time (see \citealp{Vogelsberger_2020} for a review, and as examples: Illustris -- \url{www.illustris-project.org}; EAGLE -- \citealp{Schaye_2015, Crain_2015}; and Simba -- \citealp{Dave_2019}). However, so far, the possible connection between ram pressure and SMBH accretion has been mostly studied in cosmological simulations of selected systems. For example, Romulus-C is a high-resolution, cosmological simulation of a galaxy cluster \citep[Romulus --][]{Tremmel_2017}, with which \citet{ricarte_link_2020} show that accreting SMBHs are more likely to be found in massive satellites that experience ram pressure compared to non-ram-pressure stripped galaxies. They conclude that ram pressure triggers AGN accretion and, together with SMBH feedback, can quench star formation.

The IllustrisTNG (``The Next Generation'') collaboration has developed state-of-the-art cosmological simulations \citep[][and references therein and below]{Nelson_2019_public} that naturally realize galaxies in a diversity of cosmic environments and epochs, where the SMBH feedback model has been shown to be crucial to reproducing the population of massive galaxies in the local Universe \citep{Weinberger_2018, Donnari_2019}. This could very well include jellyfish galaxies, whose frequency and physical conditions have been quantified for the first time across large cosmological samples by \citet{yun_jellyfish_2019}. The team has by now collected a sample of thousands of visually identified ram-pressure stripped jellyfish galaxies through the citizen-science effort described in \citet{zinger_jellyfish_2023}. Based on those findings, it has been possible to show that instances of jellyfish galaxies can be found also at large distances from the halo centres, in low-mass hosts and at high redshifts. The growth of gaseous tails and suppression of star formation in jellyfish galaxies are now known consequences of environmental stripping after infall into a massive host halo \citep{joshi_fate_2020, joshi_cumulative_2021, donnari_quenched_2021, donnari_quenched_2021-1}. Recently, \citet{rohr_jellyfish_2023} have quantified how ram pressure is the dominant cause of cold gas loss and can act over long timescales in jellyfish, and \citet{goller_jellyfish_2023} find bursts of star formation in jellyfish galaxies, while no enhanced star formation on the population level.

Importantly, these initial results do not preclude the important role of SMBHs, which has not been studied in detail for jellyfish galaxies in IllustrisTNG before. Among the scientific questions that still need to be tackled is the following: \textit{how are the growth and activity of SMBHs in satellite galaxies affected by the dense environments of galaxy groups and clusters?} Exploring these interactions will provide insight into the co-evolution of SMBHs and galaxies. This is the fifth paper of the team dedicated to jellyfish galaxies in the IllustrisTNG simulations. In this work, we specifically explore the presence of SMBHs in ram-pressure stripped galaxies. We describe the cosmological simulations, the SMBH model, and the selection of galaxies in Section~\ref{SEC:Methodology}. We compare the SMBH activity and AGN fraction in jellyfish galaxies to other centrals and satellites in Section~\ref{SEC:Results}, explore the possible connection between ram pressure and SMBH accretion in Section~\ref{SEC:RP_AGN}, and consider the implications of AGN feedback in Section~\ref{SEC:Discussion}. We conclude that our simulated jellyfish are hosts of more active SMBHs than other galaxies in Section~\ref{SEC:Conclusions}, likely as a result of ram pressure enhancing SMBH accretion.

%% file: sections/sec2_methods.tex

\section{Methods and Cosmological Simulations}
\label{SEC:Methodology}

We give an overview of the cosmological simulations, the SMBH model, and selection of ram-pressure stripped jellyfish galaxies here.


\subsection{SMBHs in the IllustrisTNG universe} 
                                \label{SUBSEC:IllustrisTNG}

\subsubsection{Numerical Model for Galaxy Formation and Evolution}

IllustrisTNG is a state-of-the-art suite of cosmological, gravitational + magnetohydrodynamical simulations. Originally based on the Illustris simulation \citep{Vogelsberger_2014, Genel_2014, Sijacki_2015, Torrey_2014}, it follows the formation and evolution of galaxies from the very early Universe to the current epoch, including physical processes such as star formation in the interstellar medium \citep{Springel_2003}, stellar evolution and chemical enrichment, stellar feedback driven outflows, and the growth and activity of supermassive black holes \citep{weinberger_simulating_2017} in diverse cosmic environments. The TNG galaxy formation model \citep{pillepich_simulating_2018} builds upon the original Illustris model \citep{Vogelsberger_2013} and utilizes the \textsc{arepo} code \citep{Springel_2010} to solve the coupled equations of ideal magnetohydrodynamics \citep{Pakmor_2011} and self-gravity. The details of the simulations have been widely described in the literature.\footnote{TNG Public Data Release: \url{www.tng-project.org} and \cite{Nelson_2019_public}.}

The suite consists of a series of simulation runs at three representative volumes and resolutions: TNG300 and TNG100 \citep{Springel_2018, Marinacci_2018, Pillepich_2018, Nelson_2018, Naiman_2018},  followed by TNG50 \citep{Nelson_2019, 2019_Pillepich}. In this work, we use the highest resolution runs of the last two, TNG100-1 and TNG50-1, hereafter TNG100 and TNG50. First, TNG100 has a volume of $110.7^3 \: \rm{Mpc}^3$ with
$2 \times 1820^3$ resolution elements, a baryon mass resolution of $m_{\rm{b}} = 1.4 \times 10^{6} \: \rm{M}_\odot$, and a dark matter mass resolution of $m_{\rm{DM}} = 7.5 \times 10^{6} \: \rm{M}_\odot$. In comparison, TNG50 combines both a large volume and high level of resolution in a $51.7^3 \: \rm{Mpc}^3$ box with $2 \times 2160^3$ resolution elements, a baryon mass resolution of $m_{\rm{b}} = 8.5 \times 10^{4} \: \rm{M}_\odot$, and a dark matter mass resolution of $m_{\rm{DM}} = 4.5 \times 10^{5} \: \rm{M}_\odot$. 

Halos are identified using the Friends-of-Friends (\textsc{fof}) algorithm \citep{Davis_1985}, within which gravitationally bound structures, subhalos, are identified using  \textsc{subfind} \citep{Springel_2001}. In general, subhalos with at least some stars are considered galaxies, whereby it is customary to impose a minimum stellar mass, here measured within twice the stellar half-mass radius. The main subhalo within a FoF group, which tends to be the most massive subhalo, is the central galaxy, whereas all other subhalos in the group are satellites. 

The validation against observations of the TNG simulations in modelling the cold gas content and star formation of satellite galaxies provides a robust framework for understanding ram-pressure stripping. Atomic hydrogen is preferentially removed over molecular gas from satellite galaxies \citep{Stevens_2019, Stevens_2021}, depleting their outer gas reservoirs and driving the quenching of low-mass galaxies ($M_{\rm stellar} \lesssim 10^{10}\,\rm{M}_\odot$, e.g. \citealt{Donnari_2019}) and with the quenched fractions of TNG galaxies being well within the range of observational inferences \citep{donnari_quenched_2021-1}. We can hence use the outcome of the TNG simulations to extract possible insights into how ram pressure may interact with SMBHs.

\subsubsection{SMBH seeding, growth, and feedback}

Since we are especially interested in the presence of SMBHs, we briefly summarize how black holes are seeded, how they grow, and how AGN feedback acts in the TNG simulation \citep[for full details see][]{weinberger_simulating_2017, pillepich_simulating_2018}. The TNG model places a SMBH as a sink particle with seed mass of $M_\mathrm{seed} = 8 \: \times \: 10^5 \: \mathrm{M}_\odot \:  h^{-1} \simeq 1.2\times10^{6} \: \mathrm{M}_\odot$ in a \textsc{fof} halo when it exceeds $M_\mathrm{FoF} = 5 \times 10^{10} \:  \mathrm{M}_\odot \:  \: h^{-1} \simeq 7.4 \times 10^{10} \: \mathrm{M}_\odot$. Once seeded, a SMBH grows at a rate that follows the Bondi-Hoyle-Lyttleton accretion rate $\dot{M}_\mathrm{Bondi}$, limited to the Eddington rate $\dot{M}_\mathrm{Edd}$: 
\begin{align}
    \label{eq:1}
    \dot{M}_\mathrm{Bondi} &= \frac{4 \pi G^2 M_\mathrm{SMBH}^2 \rho}{c_s^3} \\
    \label{eq:2}
    \dot{M}_\mathrm{Edd} &= \frac{4 \pi G^2 M_\mathrm{SMBH} m_p}{\epsilon_r \sigma_T c}
\end{align}
where $G$ is the universal gravitational constant, $m_p$ the proton mass, $\epsilon_r$ the radiative accretion efficiency, $\sigma_T$ the Thompson cross-section, and $c$ the vacuum speed of light. $M_\mathrm{SMBH}$ is the black hole mass, and $\rho$ and $c_s$ are the density and sound speed of the gas near the SMBH, respectively. We add that a SMBH may also grow by merging with another one during a galaxy merger event. 

As a consequence from black hole accretion, energy is released to the surrounding gas cells, which we refer to as AGN feedback. Whether the energy is released as thermal or kinetic energy depends in TNG on the Eddington ratio $f_{\rm Edd} = \dot{M}_\mathrm{Bondi} / \dot{M}_\mathrm{Edd}$, which describes how strongly a SMBH is accreting relative to its maximum possible accretion rate. In TNG, $f_{\rm Edd}$ is compared to a threshold ``$\chi$'':
\begin{align}
    \label{eq:3}
    \chi &= \min\left[\chi_0 \cdot \left(\frac{M_{\rm BH}}{10^8\, \rm M_\odot}\right)^\beta,0.1\right]
\end{align}
where $\chi_0 = 0.002$ and $\beta = 2$ are the fiducial values. Whenever the Eddington ratio is above this threshold and the accretion rate is high, AGN feedback operates via the deposit of thermal energy in the quasar mode. Conversely, when $f_\mathrm{Edd} < \chi$, energy is imparted as random kinetic kicks in the kinetic, radiatively-inefficient feedback. 

In the TNG framework, the kinetic feedback mode preventatively and ejectively quenches central galaxies, and is thought to be responsible for quenching star formation in massive galaxies \citep{Nelson_2018, zinger_ejective_2020}, including massive satellite galaxies \citep{donnari_quenched_2021, Rohr_2024}. Indeed, powerful AGN outflows can even push the gas to scales well beyond the halo boundary \citep{Ayromlou_2023a}. Likewise, we can expect AGN feedback to be important in the evolution of jellyfish galaxies.


\subsection{The ``Cosmological Jellyfish'' Zooniverse Project} 
\label{SUBSEC:Zooniverse}

The jellyfish galaxies in this study were visually classified through Zooniverse, a citizen-science platform that allows researchers to enlist the help of volunteers for identification and analysis tasks. Through the ``Cosmological Jellyfish'' Zooniverse website,\footnote{Learn more about the ``Cosmological Jellyfish'' Zooniverse Project at: \\ \url{www.zooniverse.org/projects/apillepich/cosmological-jellyfish}} the TNG team released the images for 80,704 candidate galaxies: satellites with a non-negligible gas fraction above a minimum limit $f_{\rm gas} = M_{\rm gas, \: total} / M_{\rm stellar, \: 2 \: \text{half-mass}} \geq 0.01$ and a stellar mass above $M_{\rm stellar, \: 2 \: \text{half-mass}} \geq 10^{8.3} \: \rm{M}_\odot$ for TNG50 and above $10^{9.5} \: \rm{M}_\odot$ for TNG100, between redshifts $z = 0 - 2$. The images consisted of maps of the 2-D gas mass surface density and stellar mass surface density contours in a random projection, and for each, twenty volunteers were asked to say whether the galaxy looked like a jellyfish or not.

At the conclusion of this citizen science effort, each image was assigned a score that was weighted based on the experience of the inspector, i.e. how many classifications had they already made, and accuracy, taking into account the votes of experts and the majority. A final score between 0 and 1 was given to each galaxy and, as in earlier studies, we chose a threshold of $\ge0.8$ to define a jellyfish. As a result, the TNG team collected a total sample 5,307 visually-identified jellyfish galaxies across all redshifts (and 328 jellyfish at $z = 0$), which is described in detail in \cite{zinger_jellyfish_2023}. The team also tested whether looking at a galaxy from a random orientation versus an optimized orientation significantly affected the score. Around a third of ``true'' jellyfish galaxies go undetected due to viewing angle \citep[also found in][]{yun_jellyfish_2019}. However, we respect the random orientations in our selection since they are authentic to observations.


\subsection{Selection of jellyfish galaxies and comparison samples} 
\label{SUBSEC:Sample_Selection}

Not only would we like to analyse SMBHs in jellyfish galaxies, but also compare them to SMBHs in other environments, and in galaxies that do not exhibit ram-pressure stripped tails. Our selection method is based on that of \citet{rohr_jellyfish_2023} and \citet{goller_jellyfish_2023}.  

At the first-level of selection, we consider all galaxies (hereafter black in all figures) in TNG50 and TNG100 at redshift $z = 0$ with a stellar mass $M_{\rm stellar} > 10^{9.5} \: \rm{M}_\odot$ within a host halo of mass $M_{\rm 200 c} \simeq 10^{11.5 - 14.5} \: \rm{M}_\odot$. The quantity $M_{\rm 200 c}$ represents the total mass of all matter components within a sphere whose mean density is 200 times the critical density of the Universe at the time of interest.

We distinguish between central (purple) and satellite (blue) galaxies, where the central galaxy is the most massive galaxy in the \textsc{fof} halo. For satellite galaxies that are likely to be affected by ram pressure, we look into the ``Cosmological Jellyfish'' project \citep{zinger_jellyfish_2023}. Inspected  satellites (pink) fall under the classification of satellites, but still contain some gas at the time of inspection (i.e. $f_{\rm gas} > 0.01$): these are the galaxies that were looked at to find jellyfish-like objects. Finally, jellyfish galaxies (in yellow throughout the plots) are a subset of the inspected ones, are confirmed via visual inspection to have ram-pressure stripped tails, and are selected here if they have a jellyfish score of $\geq 0.8$. Table~\ref{TAB:Sample_Selection} and Figure~\ref{FIG:Sample_Selection} give a schematic of all categories of galaxies in our sample.

\begin{figure*}
    \centering
    \includegraphics[width=0.90\linewidth]{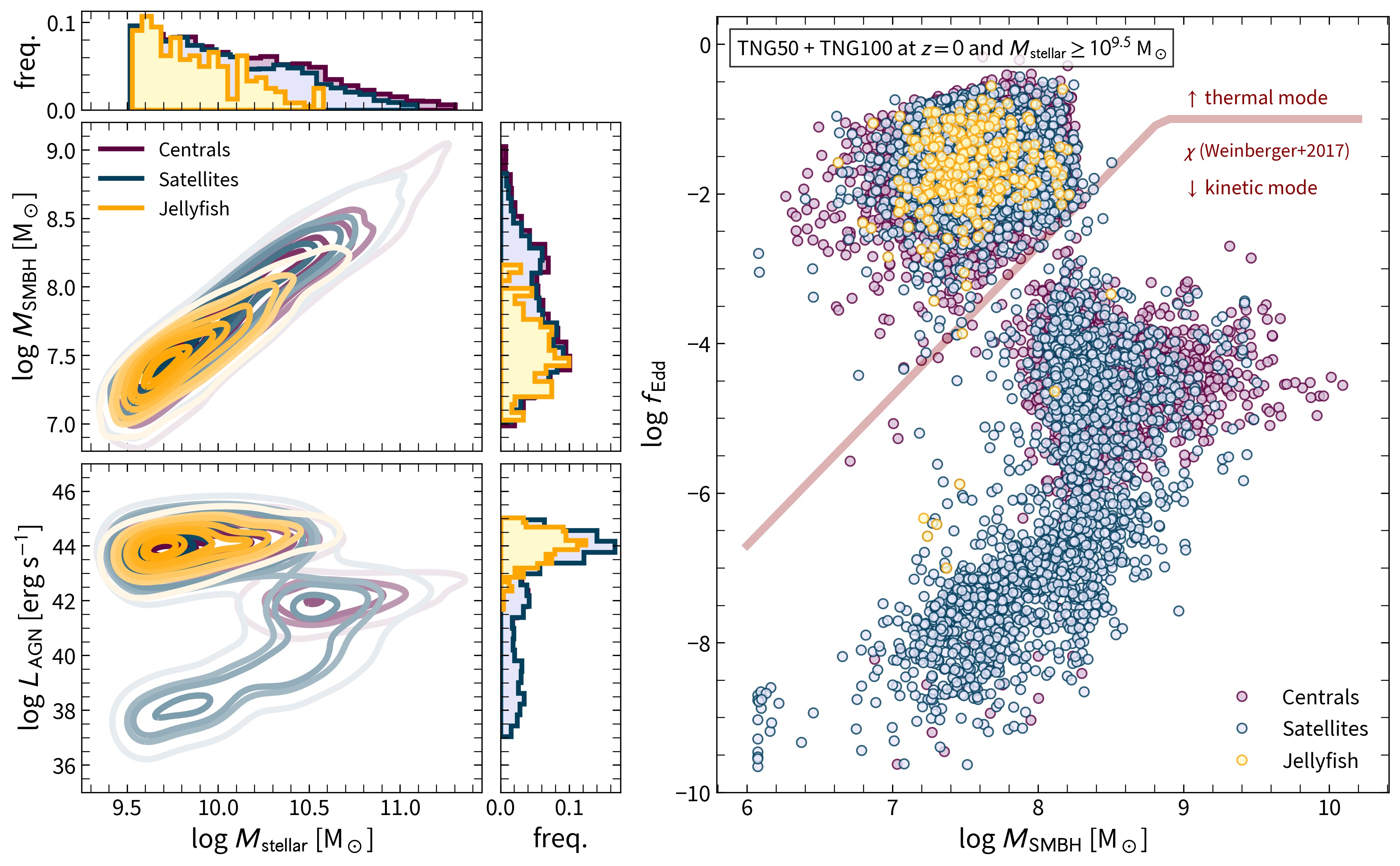}
    \caption{\textit{\textbf{Distribution of SMBH properties}} for centrals (purple), satellites (blue), and jellyfish (yellow) galaxies in TNG50 and TNG100 at $z=0$ with $M_{\rm stellar} \geq 10^{9.5} \: \rm{M}_\odot$. Operational definitions of these samples and their relative size are given in Figure~\ref{FIG:Sample_Selection}, Table~\ref{TAB:Sample_Selection}, and Section~\ref{SUBSEC:Sample_Selection}. On the left, we show the SMBH mass (top) and the AGN luminosity (bottom) as functions of the galaxy stellar mass using contours, and provide the frequency histograms for each axis normalized by the total number of galaxies, in the side panels. Each contour level represents the 10, 25, 33, 50, 67, 75, 90, 95, and 100$^\mathrm{th}$ percentiles in decreasing strength. On the right, we show the Eddington ratio as a function of $M_{\rm{SMBH}}$: according to the TNG model, galaxies with high $f_{\rm Edd}$ (above the red line) host SMBHs that exercise feedback in thermal mode, while the SMBHs of the galaxies with low $f_{\rm Edd}$ (below the red line) are in the kinetic-mode feedback \protect\cite[]{weinberger_simulating_2017}.  As a reminder, throughout this work the AGN luminosity is estimated based on the SMBH accretion rate in the simulations. Compared to centrals and satellites, jellyfish tend toward the low end of stellar and SMBH masses, but the high end of Eddington fractions and AGN luminosities.}
    \label{FIG:SMBH_Dists}
\end{figure*}


\subsection{Measurements of SMBHs and environmental effects} 
                                    \label{SUBSEC:Galaxy_Properties}

Throughout this paper, we calculate various SMBH and gas properties of the galaxies in order to measure and analyse their SMBH activity. We have already defined a galaxy's stellar mass ($M_{\rm stellar}$) as the total mass of the gravitationally-bound star particles within twice the stellar half-mass radius as a proxy for the size of the galaxy.


\subsubsection{The SMBH and AGN} 
\label{SUBSUBSEC:SMBH}

For each galaxy, we identify the SMBH as the SMBH particle that is nearest to the gravitational potential minimum, which tends to be the most massive SMBH within the galaxy, with mass $M_{\rm SMBH}$. We consider a galaxy to host an AGN if its central SMBH has a bolometric luminosity of $L_{\rm AGN} > 10^{44} \: \rm{erg \: s}^{-1}$. The calculation of $L_{\rm AGN}$ follows that used in \cite{Weinberger_2018} and \cite{Habouzit_2019}, where the radiative efficiency is $\epsilon_r = 0.2$:
\begin{equation}
    L_{\rm AGN} = \begin{cases}
                  \epsilon_r \dot{M}_{\rm Bondi} c^2 \: &, \: \dot{M}_{\rm Bondi} \geq 0.1 \dot{M}_{\rm Edd}
                  \\
                  \left(\,10 f_{\rm Edd}\,\right)^2 \,\epsilon_r \dot{M}_{\rm Bondi} c^2 \: &, \: \dot{M}_{\rm Bondi} < 0.1 \dot{M}_{\rm Edd}
                  \end{cases}
\end{equation}
Later, we also measure the kinetic energy released by the AGN, $\dot{E}_{\rm inj, \: kin}$. In the TNG outputs, this energy is given as a cumulative value over time. We calculate the $\dot{E}_{\rm inj, \: kin}$ as the difference in kinetic energy injected divided by the time between snapshots.

\subsubsection{The case of galaxies without SMBHs} 
A small fraction ($8.6$ per cent) of galaxies in our sample do not contain a central SMBH particle, even if they are massive enough at the time of inspection to host one: this is due to either the numerical seeding or the positioning methods of TNG \citep[see also][for a similar issue]{Goubert_2024}. In these cases, either a SMBH particle was never seeded in the halo or it was lost in the recent past to a more massive neighbour. The chosen value for the minimum stellar mass of the galaxies studied in this paper ($M_{\rm stellar} > 10^{9.5} \: \rm{M}_\odot$) largely alleviates this issue, as nearly all galaxies above this stellar mass do have a SMBH. We assign to the remaining few galaxies without black hole a SMBH mass and accretion rate based on a Gaussian distribution of similar-mass objects in order to keep our sample as complete and pure as possible. See Appendix~\ref{A:Missing_BHs} for more details.

\begin{table}
    \centering
    \caption{\textbf{\textit{Number of objects in our galaxy samples}} from the TNG50 and TNG100 simulations with $M_\mathrm{stellar} \geq 10^{9.5} \: \mathrm{M}_\odot$ at $\bm{z = 0.0}$. We consider `Galaxies' (black as shown), `Centrals' (purple) versus `Satellites' (blue), `Inspected' satellites (pink), and, among the latter, `Jellyfish' (yellow) galaxies. This also allows us to contrast jellyfish galaxies to the subset of inspected satellites that are not identified as jellyfish (`Inspected -- non-Jellyfish').}
    \begin{tabular}{l l l l c}                    
        \hline
        \textbf{Sample at $\bm{z=0}$}     & \textbf{TNG50}    &  \textbf{TNG100}  &   \textbf{Total}   &   \textbf{Color Key}\\
        \hline                                                
        \textit{\textbf{Galaxies}}     & 1,466             & 10,680            &   12,146           &   \swatch{black}      \\
        \textit{\textbf{Centrals}}     & 899               & 6,352             &   7,521            &   \swatch{violet}      \\
        \textit{\textbf{Satellites}}   & 563               & 4,316             &   4,879            &   \swatch{RoyalBlue}      \\
        \textit{\textbf{Inspected}}    & 410               & 2,780             &   3,190            &   \swatch{Rhodamine}      \\
        \textit{\textbf{Jellyfish}}    & 32                & 210               &   242              &   \swatch{Goldenrod}       \\
        \hline
    \end{tabular}
    \label{TAB:Sample_Selection}
\end{table}

\subsubsection{Gas near the galaxy center} 
\label{SUBSUBSEC:Inner_Gas}

The gas properties around the central region of the galaxy determine the growth and accretion of the SMBH, modulo mergers. Therefore, in the following, we study the properties of the gas mass, $M_{\rm gas}$, within $r \leq 1 \: \rm{ckpc}$ of the central black hole to gain insights on the activity of the latter. 
The number density of the gas, $n_{\rm gas}$, is the gas mass density within this radius divided by the proton mass $m_{\rm p}$, assuming that all the gas is made of atomic Hydrogen. We then calculate the pressure and sound speed of the gas as a function of temperature:
\begin{equation}
    c_s = \sqrt{\gamma \frac{P_{\rm gas}}{\mu m_{\rm p} n_{\rm gas}}}
        = \sqrt{\gamma \frac{k_{\rm B} T}{\mu m_{\rm p}}}
\end{equation}
where $\gamma = 5/3$ for a mono-atomic gas, $k_{\rm B}$ is the Boltzmann constant, and $\mu \simeq 0.59$ is the mean molecular weight. We set a temperature floor of $T = 10^3$ K to the star-forming gas cells in the calculation of $c_s$. Following \citet{yun_jellyfish_2019}, we use the mass-weighted mean values of pressure and sound speed in the chosen volume. For galaxies that do not have any gas cells within this aperture, due to the resolution or the effects of feedback, we assign them a lower gas mass, pressure, and density limit in the figures to represent them. 

In order to quantify the inflow of gas toward the SMBH, we additionally calculate the net mass inflow rate $\dot{M}_\mathrm{in - out}$. While it would be more common to compute the mass flux using a differential calculation through a thin shell around the central boundary of 1 ckpc, we resort to considering all gas cells in a slightly larger region of $R = 2$ ckpc of the SMBH to ensure sufficient sampling. Gas cells within this region that have a positive radial velocity are considered as inflowing, and conversely, cells that have a negative radial velocity are outflowing. The mass flow rate is then the product of the gas mass with the gas radial velocity, divided by the radius of the region. The net mass inflow rate $\dot{M}_\mathrm{in - out}$ is the mass outflow rate $|\dot{M}_\mathrm{out}|$ subtracted from the mass inflow rate $|\dot{M}_\mathrm{in}|$, which are specifically:
\begin{align}
    \dot{M}_\mathrm{in} &= \sum_{v_{\text{rad},i} \: > \: 0}^{\text{gas in} \: r\,\leq \,R} \frac{m_{\text{gas}, i} \: v_{\text{rad},i}}{R} \\
    \dot{M}_\mathrm{out} &= \sum_{v_{\text{rad},i} \: < \: 0}^{\text{gas in} \: r\,\leq \,R} \frac{m_{\text{gas}, i} \: v_{\text{rad},i}}{R}
\end{align}

It should be noted that the accretion rate of the SMBH, given in Equations \ref{eq:1} and \ref{eq:2} from \citet{weinberger_simulating_2017}, depends on the gas mass density $\rho$ and sound speed $c_s$ averaged over a sphere around the SMBH in a kernel-weighted fashion. While our definition of gas cells ``near'' the SMBH in this work is simpler, it is a sufficient estimate of the amount of gas that could be accreted. In addition, since we consider an aperture centred around the SMBH, the central gas quantities account for any offsets in the position of the black hole.

\subsubsection{Ambient gas around the galaxy}
                                         \label{SUBSUBSEC:Outer_Gas}

To assess the importance of the environment and ram-pressure stripping toward AGN activity, we also consider the ambient gas in the immediate vicinity of the galaxies. \cite{Ayromlou_2019} define the Local Background Environment (LBE) as an adaptive spherical shell around the galaxy and its dark matter subhalo, excluding its gravitationally bound gas. In this work, we use the mass density of the gas cells, $\rho_{\rm LBE}$, and the ram-pressure, $P_{\rm ram}$ calculated within this shell. The formula for ram pressure is \citep{Gunn_Gott_1972}:
\begin{equation}
    P_{\rm ram} = \rho_{\rm LBE, \: gas} \times v_{\rm galaxy \: rel.}^2
\end{equation}
where $v_{\rm rel}$ is the mean relative velocity of the gaseous component of a galaxy with respect to its surrounding LBE. These LBE measurements have proven to be instrumental in addressing the effects of ram pressure on galaxies across various environments, including satellites in massive systems \citep[see][]{Ayromlou_2021b}.

%% file: sections/sec3_results.tex

\section{Supermassive Black Holes in Jellyfish}         \label{SEC:Results}

In this section, we provide TNG-based predictions about the presence of AGN in jellyfish galaxies compared to other populations of satellites---which may have experienced ram-pressure stripping in the past but no longer do, or which simply experience less ram-pressure stripping---and of centrals, which do not experience environmental effects. The intention is to provide theoretical counterparts to observational findings and to ultimately assess how ram pressure, which causes the jellyfish tails, may also affect the SMBH properties.

\subsection{Jellyfish have more luminous SMBHs than other galaxies} 
\label{SUBSEC:JF_vs_Galaxies}

We start with a broad picture of the samples and the relationships among galaxy stellar mass, SMBH mass, and SMBH accretion rate. Figure~\ref{FIG:SMBH_Dists} illustrates the distribution of different properties of $z =0$ jellyfish, satellite, and central galaxies from TNG50 and TNG100. 

On the left, we show the SMBH mass and AGN luminosity as functions of the galaxy stellar mass in contours, with normalized histograms on each axis. As the overlapping contours for the samples indicate, jellyfish exhibit a similar $M_{\rm SMBH}-M_{\rm stellar}$ relation compared to satellites and centrals. However, jellyfish tend to occupy the low-end of the mass distributions, and have lower stellar and SMBH masses than satellites and centrals. Nevertheless, and importantly for the scope of this paper, almost all jellyfish (97 per cent) have AGN luminosities above $L_{\rm AGN} \gtrsim 10^{42} \: \rm{erg \: s}^{-1}$. Such a fraction is greater than for satellites (65 per cent) and centrals (85 per cent). Namely, these other subsets also include a non-negligible fraction of galaxies occupying a different locus on the $L_{\rm AGN}$ -- $M_{\rm stellar}$ plane. The separate and somewhat bimodal distribution of galaxies is an emerging feature of the TNG model and of the fact that the kinetic SMBH feedback, which is activated at relatively low SMBH accretion rates, is very effective at removing gas from the central regions of galaxies, further keeping SMBHs in the low accretion states \citep[e.g.][]{Weinberger_2018, nelson_first_2018, Pillepich_2024}.

The right-hand panel of Figure~\ref{FIG:SMBH_Dists} can provide insights as to the overall higher AGN luminosities of the jellyfish galaxies compared to the other samples. It shows a scatter plot of the SMBH accretion rate as a fraction of the maximum possible accretion rate of the black hole, the Eddington accretion rate, versus the SMBH mass. The faint red broken line delineates the separation advocated in the TNG model between the thermal and kinetic AGN feedback modes of the galaxies, as per the value $\chi$ of Equation~\ref{eq:3} from \citet{weinberger_simulating_2017}. With the exception of fewer than ten outliers (3 per cent), the SMBHs of the jellyfish accrete at high rates, and specifically at $f_{\rm Edd} > \chi$ and hence exert thermal feedback. In contrast, while a fraction of the satellites and centrals also have similarly high accretion rates, many of their SMBHs have $f_{\rm Edd} < \chi$ and are in the low-accretion, kinetic feedback mode, corresponding to lower luminosities.
 
From the very functioning of the TNG model in terms of SMBH physics, the fact that jellyfish tend to have more luminous, highly-accreting SMBHs is a result of the combination of diverse effects. First, even just to be identified as ram-pressure stripped, the jellyfish must still have some gas, and such gas can in principle accrete onto their SMBH, potentially increasing the SMBH accretion rate. Further, jellyfish have relatively smaller SMBH masses compared to the overall population, such that they have higher chances to stay above the adopted SMBH accretion rate threshold for the thermal-to-kinetic mode transition, thereby remaining in thermal SMBH feedback. To what extent these properties of jellyfish galaxies are driven by environmental versus internal processes is the question of this paper, but a few thoughts can already be laid out.

In fact, higher SMBH mass accretion rates naturally map into higher SMBH luminosities. At the same time, the SMBH feedback determines, or at least modulates, whether a black hole can accrete mass at high or low rates. In the TNG model, the thermal energy feedback is less effective than the kinetic one at reducing the gas in the SMBH vicinity, hence allowing more gas to be potentially accreted into the SMBH than in galaxies in kinetic mode. In other words, the thermal feedback mode of jellyfish seems to enable, or rather permit, them to host more luminous AGN. 

Furthermore, a face-value comparison with the other TNG galaxies allows some useful considerations. The largest majority of jellyfish have $M_{\rm stellar} \lesssim 10^{10.5} \: \rm{M}_\odot$ and $M_{\rm SMBH} \lesssim 10^{8} \: \rm{M}_\odot$, whereas the central and satellite distributions extend to greater values. As we see in Equation~\ref{eq:3}, the AGN feedback mode depends on a combination of accretion rate and mass of the SMBH and the effect is clear on the central galaxies: those that have lower stellar masses and therefore $M_{\rm SMBH} \lesssim 2-3\times10^{8} \: \rm{M}_\odot$ are in the thermal mode, while those with higher masses are in the kinetic mode and have lower accretion rates and thus less luminous AGN. 
While about 65 per cent of the satellite sample follow the same pattern as centrals, a large fraction of satellite galaxies are in the low-accretion mode despite having very small SMBHs. These satellites were once gaseous ($f_{\rm gas} \geq 0.01)$ in the past, with luminous AGN in the thermal mode, but lost much of their gas reservoirs due to environmental effects (which in turn made them transition to kinetic mode) to be inspected, much less classified as jellyfish at the time of selection \citep{zinger_ejective_2020, Terrazas_2020, donnari_quenched_2021}. We further comment on the effects of SMBH feedback on jellyfish and other galaxies in Section~\ref{SEC:Discussion}. 

\begin{figure*}
    \includegraphics[width=0.90\linewidth]{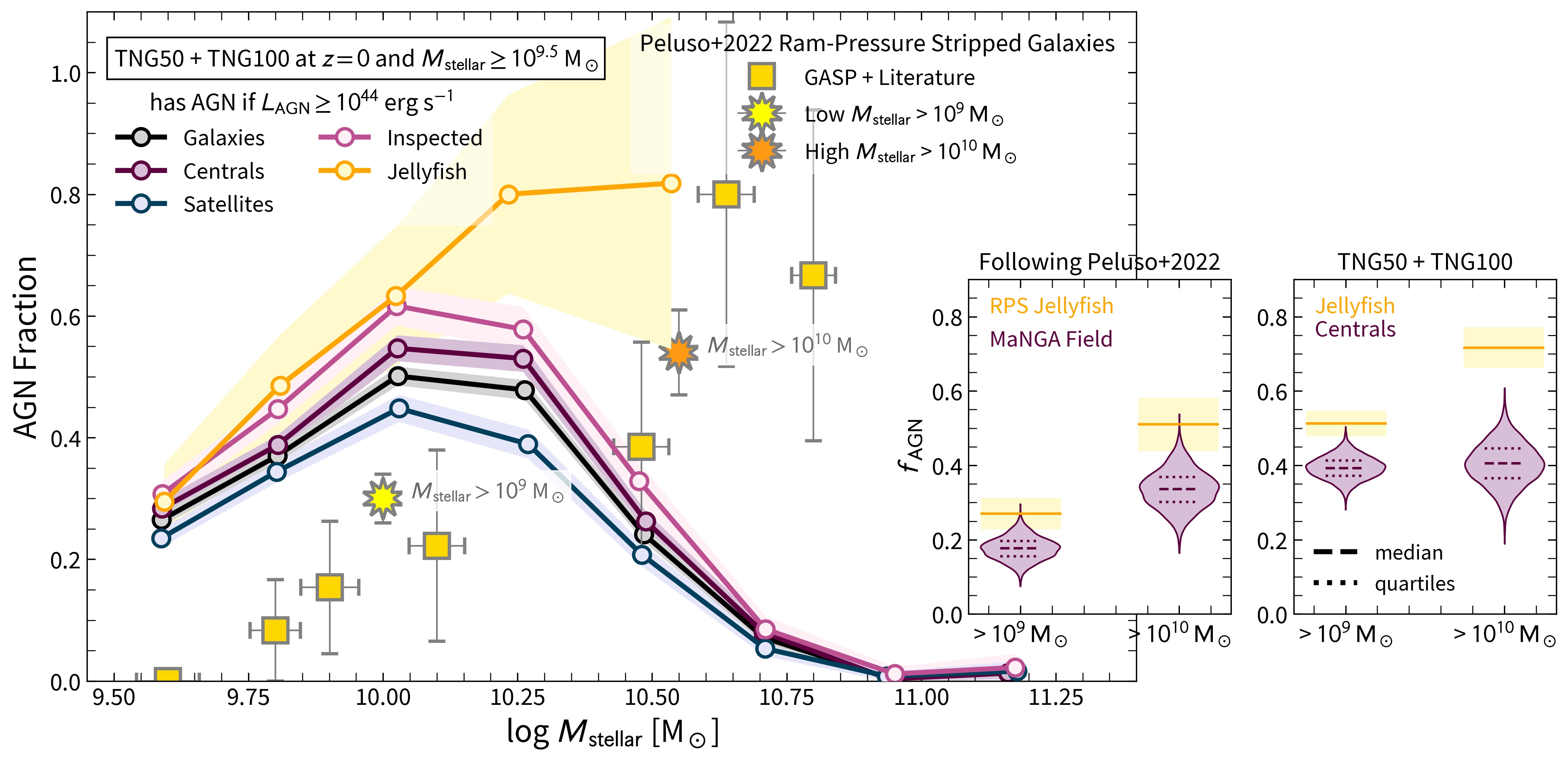}
    \caption{\textbf{\textit{The AGN fraction across various cosmological environments.}} Here we quantify the AGN fraction in stellar mass bins, i.e. the fraction of galaxies with $L_{\rm AGN} \geq 10^{44} \: \rm{erg \: s^{-1}}$, of jellyfish (yellow), compared to inspected satellites (pink), all satellites (blue), centrals (purple), and the whole population of galaxies (black) in TNG50 and TNG100 at $z=0$ with $M_{\rm stellar} \geq 10^{9.5} \: \rm{M}_\odot$. In the main panel, each bin contains at least five galaxies and is centred at the median stellar mass. We indicate the Poisson standard error of the AGN fraction in the vertical shaded regions for the simulation data. Jellyfish galaxies have higher values of $f_{\rm AGN}$ than all galaxies and satellites, as well as centrals and inspected satellites: the difference is especially strong towards higher masses, i.e. above $M_{\rm stellar} \gtrsim 10^{10} \: \rm{M}_\odot$. We juxtapose the TNG results with those for jellyfish galaxies from the GASP collaboration, who observationally derive the AGN fraction of ram-pressure stripped galaxies, binned in $M_{\rm stellar} > 10^{9} \: \rm{M}_\odot$ (light yellow star) and $M_{\rm stellar} > 10^{10} \: \rm{M}_\odot$ (dark yellow star). However, we also re-plot the \protect\citet{peluso_exploring_2022} data, calculating the AGN fraction as a function stellar mass in the same way as we do for the TNG data (yellow squares). In both those observed samples and in TNG, the fraction of AGNs in jellyfish galaxies grows with galaxy stellar mass, albeit with a different overall normalization. In the left inset panel, we emulate Figure 5 of \citet{peluso_exploring_2022} with their data: the violin plots show the probability density of $f_{\rm AGN}$ in the low and high stellar mass bins, obtained from a random sampling of the mass-matched field galaxies. The medians (dashed lines) as well as the 25 per cent and 75 per cent quantiles (dotted lines) are indicated, and compared to the AGN fraction of jellyfish galaxies (solid lines with 1-$\sigma$ shaded, in yellow). For comparison, we produce the same violin distributions with the TNG jellyfish and central galaxies on the right panel. In both observed and TNG, jellyfish are more likely to host an AGN.}
    \label{FIG:AGN_Fraction}
\end{figure*}

\subsection{More massive jellyfish have higher AGN fractions} 
                                     \label{SUBSEC:AGN_Fraction}

As discussed in the Introduction, previous observational studies, albeit not all, suggest a possible correlation between ram-pressure stripping and SMBH activity based on estimates of the AGN fraction in surveys of jellyfish galaxies. Similarly, we would like to test the incidence of AGN in the jellyfish simulated in TNG. 

The main panel of Figure~\ref{FIG:AGN_Fraction} shows the AGN fraction ($f_{\rm AGN}$) in stellar mass bins of all galaxies, centrals, satellites, inspected satellites, and jellyfish. Each bin is centred on the median stellar mass and contains at least five galaxies. In this work, we define a galaxy to host an ``active'' SMBH if its AGN luminosity is above $L_{\rm AGN} > 10^{44} \: \rm{erg \: s}^{-1}$ (see Section~\ref{SUBSUBSEC:SMBH}) since this threshold captures the high-luminosity end of the distribution in Figure~\ref{FIG:SMBH_Dists}. Compared to all galaxies and other satellites, the jellyfish consistently have higher AGN fractions at a fixed stellar mass, according to TNG. Below a stellar mass of $M_{\rm stellar} \lesssim 10^{10} \: \rm{M}_\odot$, the difference is not significant with respect to inspected satellites. However, the frequency of AGN in jellyfish continues to increase toward higher stellar masses while it declines for the other samples, although there are fewer jellyfish (41 with $M_{\rm stellar} > 10^{10.25} \: \rm{M}_\odot$, the last two bins of the plot) than other objects at such high stellar masses. In Appendix~\ref{A:L_AGN}, we test and show the outcome of different AGN luminosity cuts and find that, for all thresholds above $L_{\rm AGN} \geq 10^{41} \rm{erg \: s}^{-1}$, jellyfish consistently have higher AGN fractions than other centrals and satellites. This trend is the clearest using the fiducial choice of $L_{\rm AGN} > 10^{44} \: \rm{erg \: s}^{-1}$.

\begin{figure*}
    \centering
    \includegraphics[width=0.90\linewidth]{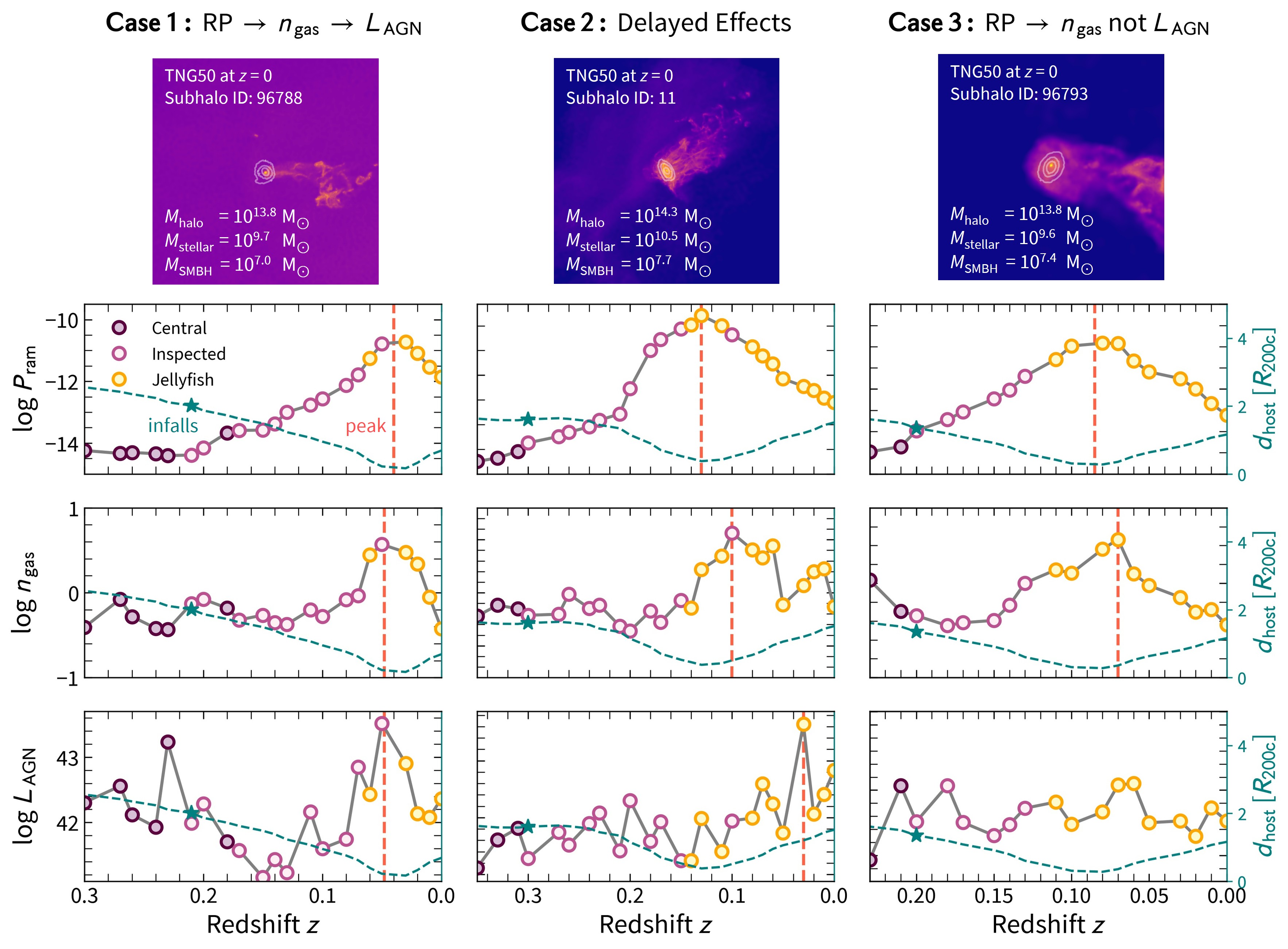}
    \caption{\textbf{\textit{Ram-pressure feeding of SMBH activity,}} shown through the evolution of ram pressure, central gas densities (innermost 1 ckpc), and AGN luminosities in three individual jellyfish galaxies from TNG50, shown in the images above at $z = 0$. We also note whether the galaxy is a jellyfish (yellow circle), inspected satellite (pink circle), satellite (blue circle), or central (purple circle) at each snapshot, as well as when the galaxy infalls into its halo (star; tracked by the host-centric distance in green). In all cases, most of the depicted quantities peak near in time to the first pericentric passage, where the host-centric distance is at a local minimum (dashed red line). \textbf{Case 1:} For \texttt{SubhaloID = 96788}, the peak in $P_{\rm ram}$ directly corresponds to a peak in $n_{\rm gas}$ and $L_{\rm AGN}$ at around the time of the first pericentric passage. \textbf{Case 2:} For \texttt{SubhaloID = 11}, the peaks in $n_{\rm gas}$ and $L_{\rm AGN}$ are slightly delayed compared to the peak in $P_{\rm ram}$, showing that ram pressure can indirectly effect SMBH activity after the pericentric passage. \textbf{Case 3:} For \texttt{SubhaloID = 97793}, the peak in $P_{\rm ram}$ corresponds to a peak in $n_{\rm gas}$ but not $L_{\rm AGN}$; here, it is possible for ram pressure to drive gas toward the centre of the jellyfish without affecting the SMBH.}
    \label{FIG:RP-AGN}
\end{figure*}

We compare the simulated AGN fraction to observations of jellyfish galaxies from two works of the GASP program, introduced in Section~\ref{SEC:Introduction}: we mark the corresponding observed AGN fractions of jellyfish galaxies with yellow squares and stars in Figure~\ref{FIG:AGN_Fraction}. The first study, \citet{poggianti_ram-pressure_2017}, finds that six out of seven jellyfish (85 per cent) host an active nucleus. These galaxies have stellar masses above $M_{\rm stellar} \gtrsim 10^{10.5} \: \rm{M}_\odot$. The second observational study, \citet{peluso_exploring_2022}, extends the search for AGN signatures to a much larger sample of 115 ram-pressure stripped galaxies from GASP and the literature, which includes X-ray and radio-selected AGN. It finds an AGN fraction of $\sim$30 per cent for low-mass jellyfish with $M_{\rm stellar} \geq 10^{9} \: \rm{M}_\odot$, and $\sim$54 per cent for $M_{\rm stellar} \geq 10^{10} \: \rm{M}_\odot$ jellyfish. In this way, both GASP and TNG return an increasing mass-dependence for the incidence of AGN activity in jellyfish galaxies. 

However, the emission-line diagnostic for identifying active nuclei in observations yields lower AGN fractions than we see in the TNG50 and TNG100 simulations. According to our constant luminosity threshold, nearly all ($\sim$$50 - 70$ per cent) of the jellyfish in TNG are hosts of AGN, which is about 20 percentage points larger than in the aforementioned observational samples. It should be kept in mind that the observed galaxies stem from a compilation of studies across wavelengths and selection functions, making direct, quantitative comparisons with simulations difficult. While raising the AGN luminosity threshold for TNG galaxies may lower the values of $f_{\rm{AGN}}$ to match with observations, our fiducial cut of $L_{\rm{AGN}} \geq 10^{44} \: \rm{erg \: s}^{-1}$ is already a few orders of magnitude higher than the X-ray luminosities reported by those studies \citep[$L_{\rm{AGN}} \simeq 10^{40 - 42} \: \rm{erg \: s}^{-1}$;][]{peluso_exploring_2022, Tiwari_2024, Koulouridis_2024}. Moreover, we do not make mock observations of the AGN luminosity, and note that a theoretical criteria based on the Bondi accretion rate in a subgrid framework may not necessarily return realistic outcomes in the details. These underlying differences in our characterization of the AGN might also contribute to the discrepancy in the slope.

Still, the outcome of our analysis shows consistency between observations and simulations: jellyfish have higher incidences of AGN activity than central and other satellite galaxies. The violin plots in the sub-panels of Figure~\ref{FIG:AGN_Fraction} further demonstrate this result in the same style of Figure 5 by \citet{peluso_exploring_2022}. On the left sub panel, we reproduce the results by the GASP programme, with their observed data. The figures show the probability distribution of $f_{\rm AGN}$ values for central galaxies, compared with the $f_{\rm AGN}$ of jellyfish galaxies, split in low and high stellar-mass bins. On the right sub-panel, we construct the probability distributions from bootstrap random extractions of the TNG central galaxies with ten-thousand subsamples, each matching the number of galaxies in stellar mass bins of $\sim0.3$ dex. \citet{peluso_exploring_2022} compare their jellyfish sample with a Mapping Nearby Galaxies at Apache Point Observatory (MaNGA) reference sample of field galaxies that are not affected by ram-pressure stripping according to halo mass estimates and visual inspection. 
Similar to the results by \citet{peluso_exploring_2022}, TNG jellyfish galaxies are more likely to host an AGN than non-ram-pressure stripped galaxies, albeit with overall higher AGN fractions and with the difference being less significant toward lower stellar masses than in the observed samples.

To summarize, TNG jellyfish tend to host SMBHs that accrete at higher rates and hence have higher AGN occupation fractions than centrals and other satellites, especially for high-mass galaxies above $M_{\rm stellar} \gtrsim 10^{10}\, {\rm M_\odot}$. They are located toward the lower-mass end of the $M_{\rm SMBH}-M_{\rm stellar}$ relation. As a result, their SMBHs are still in the thermal feedback mode, have high accretion rates, and have overall higher luminosities than centrals and satellites. This leads to higher incidences of AGN in jellyfish compared to the other galaxy samples. These findings suggest that environmental interactions, such as ram-pressure stripping, may play a role in fuelling SMBH growth and sustaining AGN activity in jellyfish galaxies. We remind that jellyfish and inspected satellites are distinct from the satellite population with respect to their gas fractions, in that ram-pressure stripped galaxies in the ``Cosmological Zooniverse'' sample must have $f_{\: \rm gas} > 0.01$. In general, on the other hand, the $z=0$ satellite population is mostly gas-poor, which has been shown in both the TNG simulations and in observations \citep[e.g.][]{Stevens_2019, Stevens_2021}. If a satellite is gas-poor, especially in its central regions, then it almost certainly has a low SMBH accretion rate and would not be considered to have an AGN. Even so, as suggested by observational studies, these results might point to a connection between dense cosmological environments and SMBHs. However, we need to further examine the gas properties of the jellyfish to make further conclusions on the role of ram pressure.

To this point, we have addressed the presence of SMBHs and their properties in jellyfish galaxies versus in all galaxies, centrals, and satellites. As some observational studies suggest, a possible explanation for the higher AGN luminosities of the jellyfish galaxies than all other galaxies is that the higher density and ram pressure of their cosmological environment can drive more gas toward the galactic nuclei, increasing the supply of gas to the SMBH. In this way, ram pressure might be able to feed AGN activity.

\section{A Ram Pressure and AGN Connection?}
\label{SEC:RP_AGN}

Irrespective of cosmological environments, a SMBH accretion rate depends not only on its mass but also on the physical properties of the gas in its immediate surroundings. In particular, in the case of the Bondi formulation (Equation~\ref{eq:1}), a SMBH accretion rate is directly proportional to the density of the neighbouring gas via $\rho$, in addition to being inversely proportional to its sound speed via the $c_s^3$ factor. Therefore, the higher the density of the gas in the central regions of galaxies, the more luminous are their AGN. In turn, the availability and hence density of the gas at the centre of galaxies may depend on both secular processes---such as star formation and SMBH feedback---as well as phenomena driven by the cosmological environments, such as ram pressure.

Our findings so far include a static snapshot in time and, while the simulations provide a large sample to study, they also offer the advantage of following galaxies across time. At $z\simeq0$, we are actually probing jellyfish over many evolutionary states: some are just infalling into their host, while others have been in orbit for billions of years. A more informative analysis may be to look at the evolution of galaxies from being centrals to becoming jellyfish, in order to determine if and how the AGN luminosity changes with ram pressure.

In the following, we investigate the possible connection between ram pressure and AGN and to understand \textit{how} a process that occurs outside a galaxy could be tied to SMBH accretion, which occurs at the very centre of galaxies. We proceed in an outside-in fashion, by connecting ram pressure to central gas density and hence to AGN luminosity with the goal of addressing why jellyfish exhibit higher AGN luminosities than, not only central galaxies, but also all other satellites. We then return to the fundamental role of secular internal processes, and chiefly of SMBH feedback, in the Discussion.

\subsection{Studies of ram pressure and SMBH evolution}

We perform a case-by-case analysis of the TNG50 jellyfish sample where we follow the evolution of the ram pressure, central gas density, and AGN luminosity since before their infall, through their pericentric passages, and to the present-day. As a reminder, the ram pressure is measured from the portion of the intragroup or intracluster medium that is located in the surroundings of the galaxy, whereas the central gas density is with $r \leq 1$ ckpc of the SMBH (see Section~\ref{SEC:Methodology}. The time of infall $\tau_{\rm infall}$ is the first simulated snapshot that a galaxy becomes a satellite member of its $z = 0$ host halo. The time of pericentric passage $t_{\rm pp}$ is when the host-centric distance is at a local minimum. By inspecting the evolutionary tracks of 242 jellyfish galaxies from TNG50 and TNG100, we identify three classes of systems, which we showcase via three representative cases in Figure~\ref{FIG:RP-AGN}. For each case, we show the image of a jellyfish galaxy at $z = 0$ and track its $P_\mathrm{ram}$, $n_\mathrm{gas}$, and $L_\mathrm{AGN}$ over time. \\

\noindent\textit{\textbf{Case 1: \textit{Direct} Effect of Ram Pressure $\rightarrow$ Central Gas $\rightarrow$ AGN}} \\

In the left column of Figure~\ref{FIG:RP-AGN}, the ram pressure $P_{\rm ram}$ increases as the host-centric distance $d_{\rm host}$ decreases, and peaks at the time of the first pericentric passage along with $n_{\rm gas}$ and $L_{\rm AGN}$. Higher ram pressure directly correlates with higher central gas densities, driving SMBH accretion and enhancing the AGN luminosity. Approximately 50~per~cent of jellyfish show signs of elevated central gas density and AGN luminosity simultaneously with ram pressure. \\

\noindent  \textit{\textbf{Case 2: \textit{Delayed} Effect of Ram-Pressure $\rightarrow$ Central Gas $\rightarrow$ AGN}} \\

In the middle column  of Figure~\ref{FIG:RP-AGN}, $P_{\rm ram}$ once again peaks at the time of the first pericentric passage, however $n_{\rm gas}$ peaks slightly later, and there is a delayed rise in $L_{\rm AGN}$. Here, ram pressure seems to indirectly affect SMBH activity some time after the pericentric passage. Approximately 15~per~cent of jellyfish show delayed signs of elevated central gas density and AGN luminosity. \\

\noindent\textit{\textbf{Case 3: Effect of Ram-Pressure $\rightarrow$ Central Gas \textit{but not on} AGN}} \\

Finally, in the right column of Figure~\ref{FIG:RP-AGN}, the peak in $P_{\rm ram}$ corresponds to a peak in $n_{\rm gas}$ at the time of the first pericentric passage. Higher ram pressure still seems to directly increase the central gas density, but does not affect the AGN luminosity. Approximately 35~per~cent of jellyfish show signs of elevated central gas densities but not of AGN luminosity during the peak in ram pressure. \\

\begin{figure}
    \centering
    \includegraphics[width=0.93\linewidth]{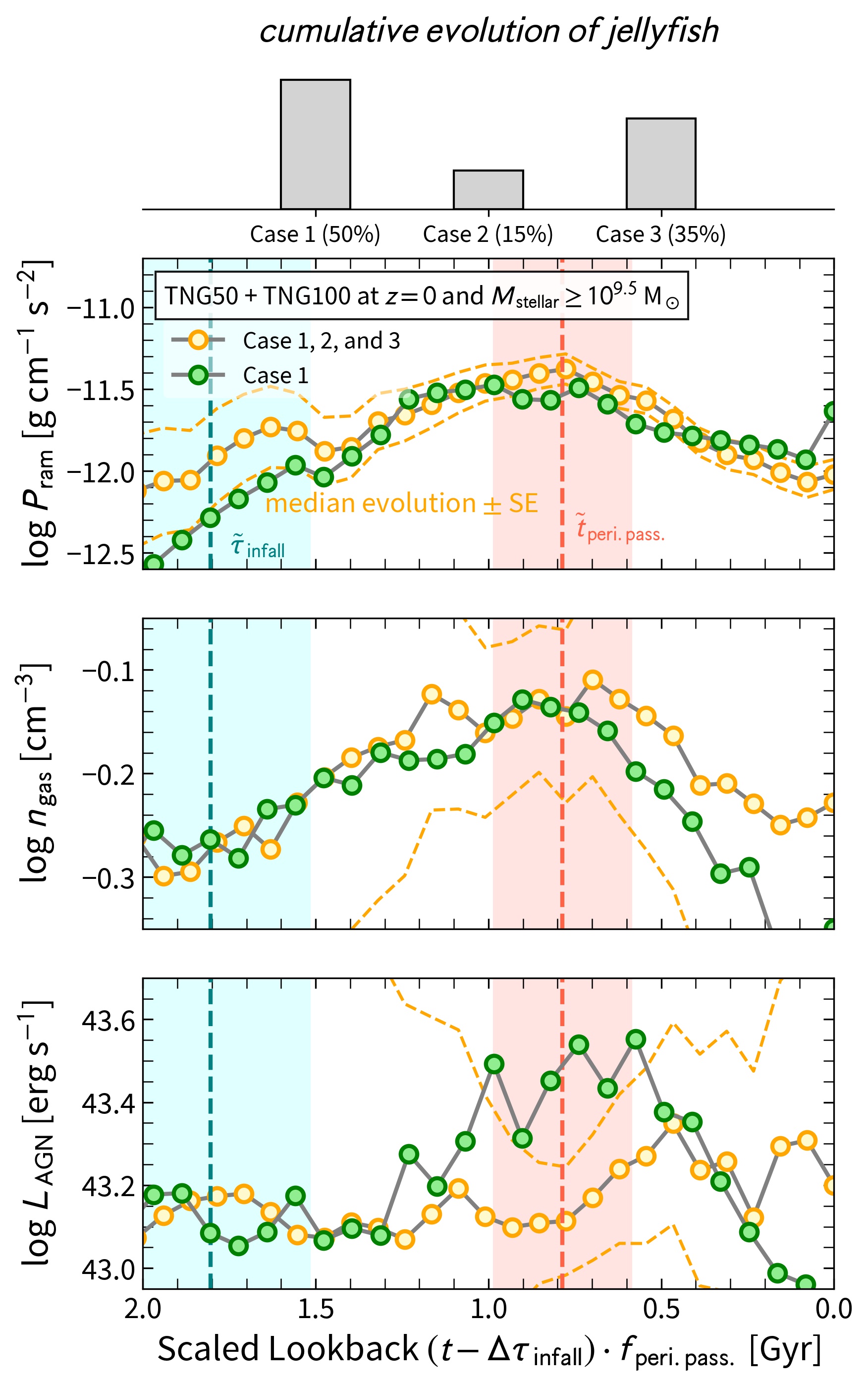}
    \caption{\textbf{\textit{Ram-pressure feeding of SMBH activity,}} shown through the evolution of ram pressure, central gas densities, and AGN luminosities in hundreds of jellyfish galaxies from TNG50 and TNG100 at $z = 0$. The cumulative i.e. ``stacked'' evolution represents a combination of three evolutionary paths by which ram pressure affects the AGN activity of the jellyfish galaxies: we categorize them as ``Case 1'' (Direct Effect), ``Case 2'' (Delayed Effect), and ``Case 3'' (No Effect) and quantify them in the above bar plot by visual inspection of the evolutionary tracks of all individual galaxies (as those of Figure~\ref{FIG:RP-AGN}). We show the median-stacked evolution, with the galaxy-to-galaxy standard deviation around the median as the dashed outer lines. The standard error extends beyond the figure limits in the case of $n_{\rm gas}$ and $L_{\rm AGN}$, as these quantities are highly variable between galaxies. To stack the data, we shift the time axes by $\Delta \tau_{\rm infall}$ and scale by $f_{\rm peri. \: pass.}$ to align the evolutions at the average time of infall (dotted teal line) and first pericentric passage (dotted red line with 1-$\sigma$ shaded). The time correlation between ram pressure and AGN luminosity is somewhat weak. However, when we stack across only jellyfish galaxies of the first case (green circles), a strong and timely increase of AGN luminosity manifests with an increase of ram pressure and central gas density.
    } 
    \label{FIG:RP-AGN_Cumulative}
\end{figure}

A ``cumulative'' time evolution represents the most common or average fate of the jellyfish galaxies. We show this in Figure~\ref{FIG:RP-AGN_Cumulative}, where we stack the otherwise disparate evolution of hundreds of galaxies by rescaling the lookback time of the simulation snapshots to account for both the diverse times of infall and of first pericenter passage across the sample \citep[see][who use a similar method to stack the orbital evolution of satellites in EAGLE]{Wright_2022}. In fact, on average, the $z=0$ jellyfish galaxies of TNG50 and TNG100 infall $\Tilde{\tau}_{\rm infall} \simeq 4.2$ Gyr ago and experience their first pericentric passage about $\Tilde{t}_{\rm pp} \simeq 2.2$ Gyr. Late-falling jellyfish which have not had a pericentric passage before $z=0$ are included in the median. We shift the evolution by $\Delta \tau = \tau_{\rm infall} - \Tilde{\tau}_{\rm infall}$ such that the stacked infall times align at the median $\Tilde{\tau}_{\rm infall}$, and then scale by a factor of $f_{\rm pp} = \Tilde{t}_{\rm pp} / (t_{\rm pp} - \Delta \tau)$ so that the stacked times of the first pericentric passages fall on the median $\Tilde{t}_{\rm pp}$. The horizontal axis of the stacked evolution thus reads as the scaled lookback time $f_{\rm pp} \cdot (t - \Delta \tau_{\rm infall})$. Then, we interpolate and average the evolutions on evenly-spaced points of time.

\begin{figure}
    \centering
    \hspace{1.5em}
    \includegraphics[width=0.899\linewidth]{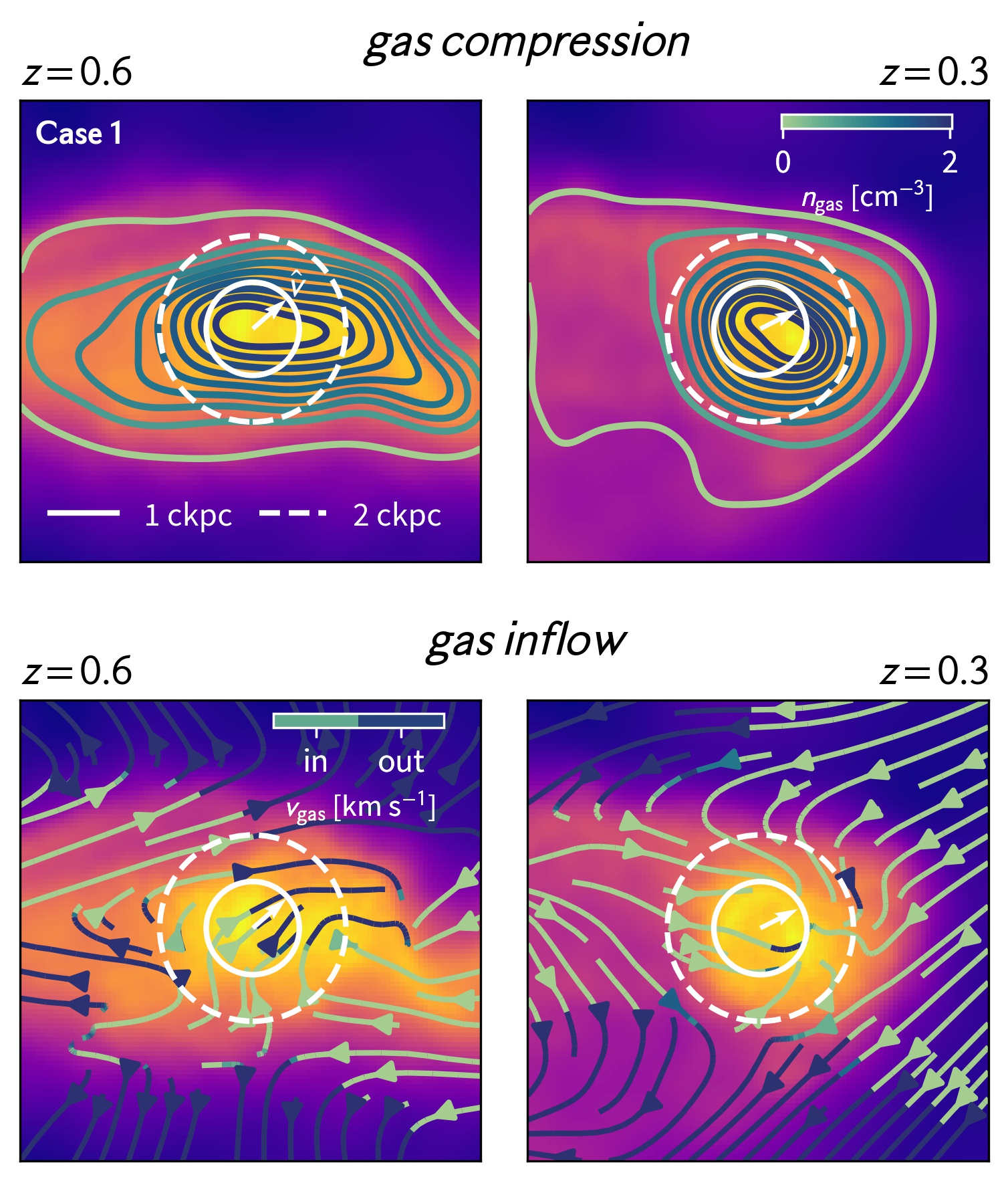}
    \caption{\textbf{\textit{Possible mechanisms of ram-pressure feeding.}} Here, we show the 2-D gas density projection of the example jellyfish galaxy in ``Case 1'' of Figure~\ref{FIG:RP-AGN}, where there is a direct effect between ram pressure, central gas density and SMBH luminosity, across two snapshots in time: shortly before ($z = 0.6$) and during ($z = 0.3$) the peak in ram pressure. The inner $r = 2$ ckpc (dashed white circle) and $r = 1$ ckpc radii (solid dashed circle) are shown as visualizations of where we calculate the net central inflow rate, where the gas could be accreted onto the SMBH. The galaxy's direction of motion is marked with its velocity vector $\hat{v}$ (white arrow). The first row shows the gas density contours: as the jellyfish experiences more ram pressure, the gas appears to be compressed into the central regions. In the second row, we consider the velocity flow of the gas: compared to the first snapshot, the gas in the second snapshot has more inflows toward the SMBH at the galaxy centre.}
    \label{FIG:Mechanisms}
\end{figure}

\begin{figure}
    \centering
    \hspace{2em}
    \includegraphics[width=0.78\linewidth]{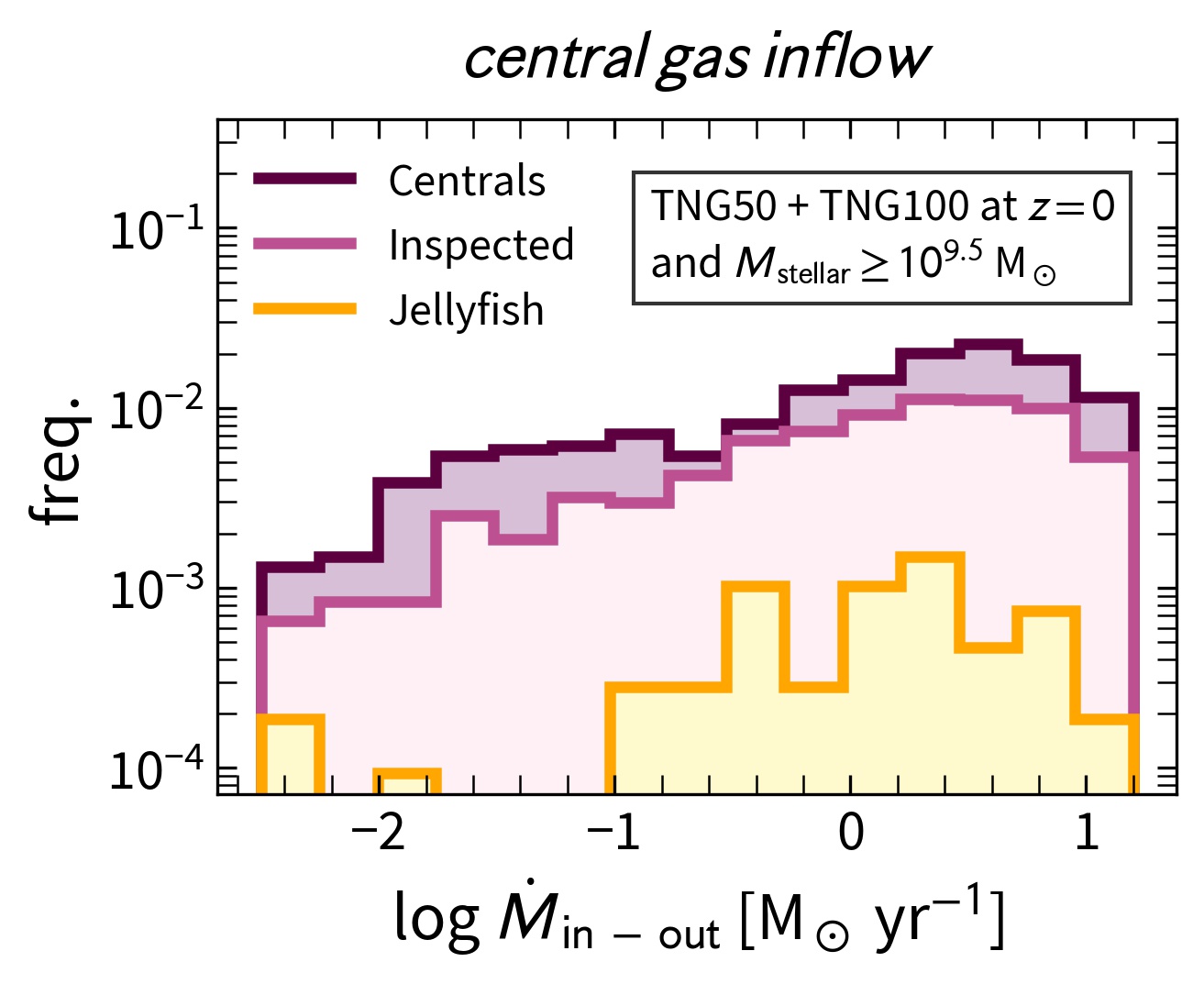}
    \contcaption{Above, we show the distribution of the net inflow rate of the central gas of the jellyfish (yellow), inspected satellites which are \textit{not} jellyfish (pink), and centrals (purple) in TNG50 and TNG100 at $z = 0$ with  $M_{\rm stellar} \geq 10^{9.5} \: \rm{M}_\odot$. Each bin represents the frequency of galaxies out of all galaxies shown. The net mass inflow rate is measured within the central 2 ckpc of the galaxy, as in the region illustrated by the solid and dashed white circles in the 2-D projections above. The jellyfish experience overall higher amounts of inflowing gas in the central regions than the other galaxies. Therefore, ram pressure likely channels and increases the density of gas toward the centre and in this way enhances SMBH accretion.}
    \label{FIG:Mechanisms_continued}
\end{figure}

The stacked evolution in Figure~\ref{FIG:RP-AGN_Cumulative} resembles the first or second case of Figure~\ref{FIG:RP-AGN}, which represent the majority, although the rise in the AGN luminosity is weak. 
The accretion rate onto SMBHs is highly variable with time: this could already be appreciated in the single cases of Figure~\ref{FIG:RP-AGN} across snapshot points separated by about 150 Myr. In fact, individual TNG galaxies have been shown to exhibit even shorter variability time scales of SMBH accretion rates when analysed with time cadence of a few million years, for example, from Milky Way-like galaxies \citep{Pillepich_2021} to massive ellipticals at the centres of clusters \citep{Prunier_2025, Lehle_2025}. This makes the evolution of the SMBH accretion rates somewhat noisy when stacking across a couple hundred galaxies. Moreover, there is variation in how the jellyfish evolve---all three scenarios are possible: even if the time tracks are stacked to account for the differences in pericentric passages, the possible enhancements of AGN luminosity across galaxies of the first or second case will necessarily be smoothed out over large time intervals, if only because different galaxies of the second case will exhibit such peaks at slightly different times. We test, but we do not show, whether the peak in the AGN luminosity would be stronger in a high stellar mass bin ($M_{\rm{stellar}} \geq 10^{10} \: \rm{M}_\odot$) since jellyfish galaxies above this mass have significantly higher AGN fractions compared to other samples of galaxies. However, we do not see a clearer connection between $P_{\rm{ram}}$, $n_{\rm{gas}}$, and $L_{\rm{AGN}}$ in the stacked evolution. We attribute this again to the high variability in the SMBH accretion rate between snapshots.

Nevertheless, when we stack across jellyfish galaxies of the first case (``Case 1'' Direct Effect, green circles in Figure~\ref{FIG:RP-AGN_Cumulative}), we do recover a stronger increase in the AGN luminosity of about $\sim 0.5$ dex when ram pressure and central gas density peak. So, according to TNG and our analysis thus far, jellyfish galaxies experience more ram pressure as they approach the centre of their host halo \citep[see also][for more details]{rohr_jellyfish_2023}. By the time of the first pericentric passage, ram pressure has peaked, along with the central gas density. The elevated central gas returns an elevated gas mass accretion onto the SMBH, and hence elevated SMBH luminosity, though this is not the case for all jellyfish galaxies. What determines whether a jellyfish experiences a direct or delayed increase in AGN luminosity with ram pressure, or no effect at all, is a complex question. The galaxy stellar mass, how long it has been infalling, or even its inclination angle as it approaches its host halo,\footnote{Wind-tunnel simulations might provide insight in this case. For example, \citet{Sparre_2024} have explored the effect of inclination angle on ram pressure stripping itself, although not the implications on the central gas.} could either inhibit or enhance the impact of ram pressure on the SMBH. For instance, the example of a ``Case 2'' jellyfish is more massive than the average jellyfish, and an order of magnitude more massive than the galaxies in the other two cases, possibly accounting for the delayed effects. However, ``Case 2'' jellyfish do not have higher stellar masses overall. We do not expect any single factor alone can differentiate all ``Case 1'' jellyfish from ``Case 2'' or ``Case 3'' at the population level, because the hundreds of jellyfish in our case studies cover a diversity of evolutionary paths.

In this way, TNG, or at least a large majority of jellyfish therein, supports the observational hypothesis that ram-pressure stripping feeds SMBH activity in jellyfish galaxies, helping to explain their higher AGN fractions. Our findings are further similar to those of \citealt{ricarte_link_2020} based on the galaxy populations of the simulated galaxy cluster Romulus-C, where SMBH accretion is enhanced during pericentric passage. In fact, our findings are also consistent with previous results based on TNG satellites that would appear, at a first glance, contradictory. For example, \citealt{joshi_fate_2020} showed that gas accretion into SMBHs is ultimately significantly suppressed for satellite galaxies after their accretion into cluster-mass hosts in comparison to galaxies of similar mass ``in the field.'' This is a long-lasting and population-averaged environmental obstruction to SMBH growth and activity that is not in contrast with short phases of enhanced AGN accretion, and thus luminosity, across the history of individual galaxies. This is indeed in line with what happens to star formation \citep{{goller_jellyfish_2023}}, with individual bursts of star formation coexisting with the ultimate quenching of the satellite population.

\begin{figure*}
    \centering
    \hspace{-2em}
    \includegraphics[width=0.9\linewidth] {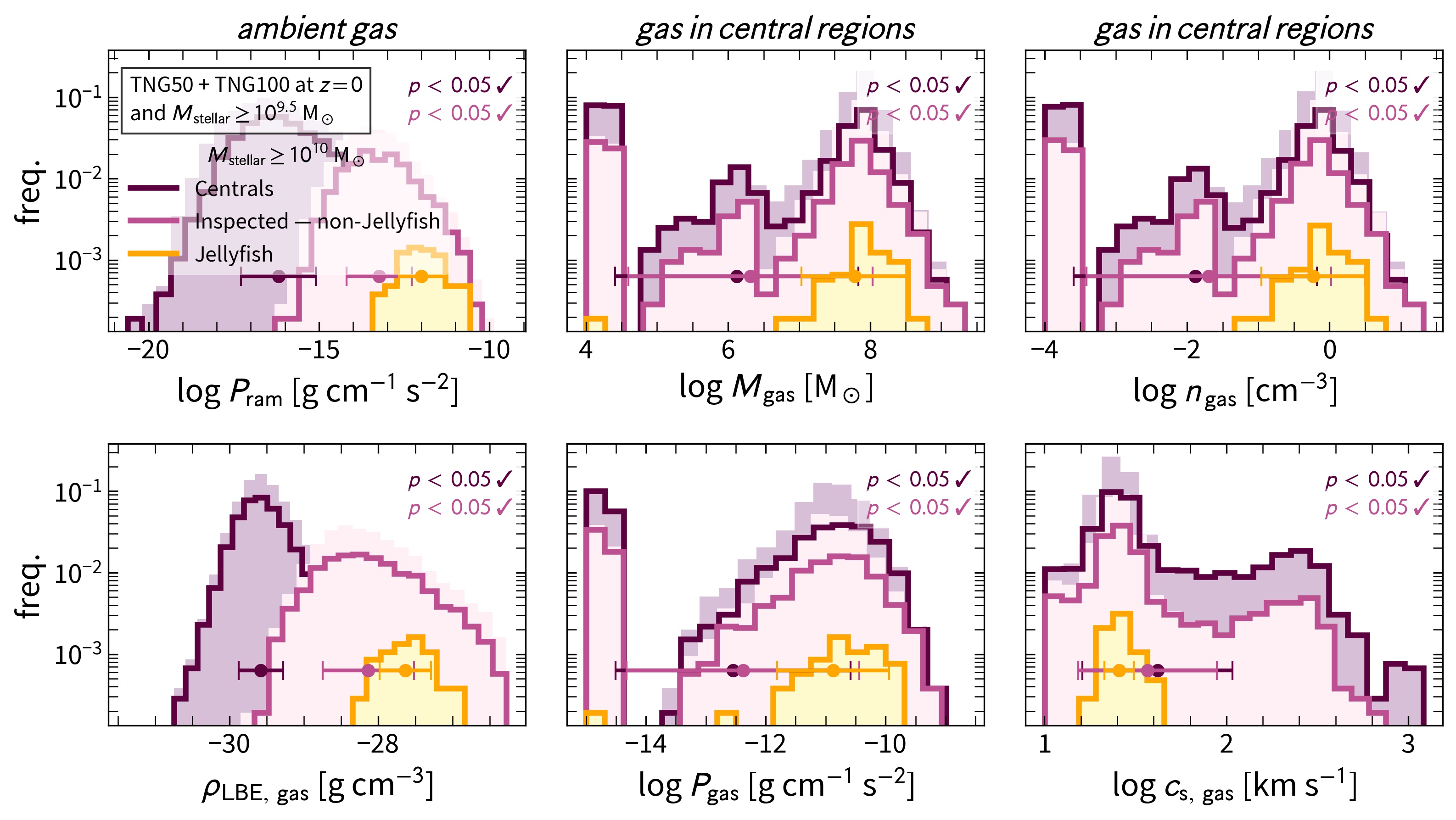}
\caption{\textbf{\textit{Distribution of gas properties in and around jellyfish, gaseous satellites, and central galaxies.}} We show the distributions of properties of the internal and the surrounding gas of the jellyfish (yellow) and inspected satellites which are \textit{not} jellyfish (pink), and centrals (purple) in TNG50 and TNG100. We focus on $z=0$ and show galaxies with $M_{\rm stellar} \geq 10^{9.5} \: \rm{M}_\odot$ (shaded histograms) and with $M_{\rm stellar} \geq 10^{10} \: \rm{M}_\odot$ (solid-line histograms). Each bin represents the frequency of galaxies out of all galaxies shown. For the more massive sub-samples, we provide the results of a two-sample Kolmogorov-Smirnov (K-S) test, at the top-right corners of each panel. In the first column, we show properties of the ambient gas, defined as the gas in the local background environment (LBE) of the galaxy: the ram pressure $P_{\rm ram}$ and density $\rho_{\rm LBE, \: gas}$, only applicable to satellites. The second and third columns show the properties of the central gas with $r \leq$ 1 ckpc of the SMBH: the central gas mass $M_{\rm gas}$, density $n_{\rm gas}$, pressure $P_{\rm gas}$ and sound speed $c_{\rm s, \: gas}$. In order to still give representation to those galaxies that do not have any gas cells within such a small radius, we assign them very low values of $M_\mathrm{gas}, P_\mathrm{gas}, n_\mathrm{gas}$, and so on, seen as the peak at the lower limits of the histograms (see Section~\ref{SUBSUBSEC:Inner_Gas}). At high stellar masses, the jellyfish tend toward the end of galaxies with high central gas masses, densities, and pressures and are statistically different from both centrals and inspected satellites, with low $p$-values.}
    \label{FIG:Gas_Dists}
\end{figure*}


\subsection{Possible mechanisms for ram pressure feeding the SMBHs}
\label{SUBSEC:Mechanisms}

As Figures~\ref{FIG:RP-AGN} and~\ref{FIG:RP-AGN_Cumulative} demonstrate, dense cosmological environments, such as the gaseous halos of groups and clusters,  have a marked influence on the internal properties of jellyfish galaxies. Across the individual and average cases, we see that the density of the gas within a kiloparsec of the SMBH increases along with ram pressure as the jellyfish approaches the halo centre, regardless of whether the effect surfaces in the evolution of the AGN luminosity. \textit{But why should ram pressure, which acts in the outskirts of jellyfish galaxies, affect the gas at their centres, near the SMBH?} Following previous works, we speculate that this occurs by enhancing gas inflows towards the galactic centre, by simply compressing existing gas or by decreasing outflows. We support this physical interpretation as follows.

In Figure~\ref{FIG:Mechanisms}, we take the example ``Case 1'' jellyfish of Figure~\ref{FIG:RP-AGN} and plot its 2-D density contours and velocity flow shortly before and during the galaxy's pericentric passage. Before the peak in ram pressure, the densities are distributed throughout the disc of the jellyfish and gas flows into (green) and out (blue streams) of its centre. During the peak, densities become concentrated in the inner regions, and there are more gas flows into the galaxy's centre from all directions, particularly at the leading edge where ram pressure occurs. If 50 per cent of the TNG jellyfish can be categorized under ``Case 1''---65 per cent including ``Case 2'' of delayed effects---then it is possible that for many of the jellyfish, ram pressure feeds the SMBH by way of gas compression and flows toward the centre. 

To explore these mechanisms further, we measure the net mass inflow rate of the central gas at the boundary of 1 ckpc (solid white circle), mass-weighted and averaged within 2 ckpc (dashed white circle), as illustrated in the same 2-D density projections, but for all galaxies in our sample. Further details on how we measure the net mass inflow rate are in Section~\ref{SUBSUBSEC:Inner_Gas}. As a matter of fact, the histogram in Figure~\ref{FIG:Mechanisms} confirms that jellyfish galaxies experience significantly more inflows of gas in the central regions compared to other satellites and centrals (more on this comparison next, in Section~\ref{SUBSEC:Linking}). Or more precisely, they are biased towards the larger values of net inflow rates towards their center than centrals and inspected satellites. This demonstrates that ram pressure can increase the density of gas near the SMBH, while decreasing the reservoir of gas in the outer regions of the galaxy. The elevated central gas density and accretion rate thus increase the AGN luminosity. 

As introduced in Section~\ref{SEC:Introduction}, observational studies have similarly suggested gas compression as a possible channel through which ram pressure causes gas to flow toward the centres of jellyfish galaxies \citep[][]{Vulcani_2018, Vulcani_2024, Roberts_2022, Roberts_2023}. The TNG-based insights of Figure~\ref{FIG:Mechanisms} also appear to be consistent with other theoretical studies of the ram pressure and AGN connection. As we introduced in Section~\ref{SEC:Introduction}, simulations with more detailed and physically-motivated implementations of the interstellar medium and SMBH accretion can shed more light on the exact mechanisms that lead to more gas at the galactic centre. For example, the wind-tunnel hydrodynamical simulations of \citet{Akerman_2023} consider a torque model for accretion, different from the Bondi formula also used by TNG, and finds that the ram pressure leads to the mixing of interstellar and intracluster gas, increasing the torque at the inner regions of the galaxy disk.  Using the standard Bondi accretion, the TNG simulations find a similar qualitative trend of increasing SMBH accretion rates with increasing ram pressure. In simulations of idealized galaxies from \citet{Schulz_2001} and \citet{Tonnesen_2009, Tonnesen_2012}, ram pressure strips away low-density gas on the outskirts while compressing and removing angular momentum from the remaining high-density clouds. The loss of momentum causes the clouds to drift and spiral toward the SMBH to be accreted. Zoom-in simulations that include magnetohydrodynamics \citep{Ramos_Martinez_2018} additionally consider the potential influence of magnetic fields: ram pressure induces shocks in a flared, magnetized disk and funnels inflows of gas toward the centre to be accreted by the SMBH.

The channels suggested by these simulation experiments, through which ram pressure feeds AGN activity, are all possible explanations for the increase of central gas densities and SMBH accretion that we see naturally accounted for in most TNG jellyfish galaxies. However, re-simulating these galaxies in a resolved multiphase interstellar medium and with a more realistic black hole accretion model would be needed to complement our results. While magnetic fields could also have a role, they are outside the scope of this work and would make an interesting avenue for future study.


\subsection{Linking the environment to the gas inside the galaxy}
\label{SUBSEC:Linking}

While we have found evidence for a direct connection between dense cosmological environments and dense galactic centres in Figure~\ref{FIG:Mechanisms}, it is important to explore if this holds across gas-rich satellites that have been identified as jellyfish in contrast with those that have not---\textit{are jellyfish really unique from other satellites?} According to TNG50 and TNG100, the link between the environment and internal gas near the SMBH is consistent at the global level and distinguishes jellyfish from centrals and other gaseous satellites.

Figure~\ref{FIG:Gas_Dists} shows the 1-D histograms of the ambient and central gas properties of central galaxies, inspected satellites that have \textit{not} been visually identified to have ram-pressure stripped tails, and jellyfish galaxies from the simulations,  with a focus on high-mass satellites with $M_{\rm stellar} \geq 10^{10} \: \rm{M}_\odot$. Above this stellar mass, jellyfish have significantly higher AGN fractions than all other categories of galaxies (see Figure~\ref{FIG:AGN_Fraction} in Section~\ref{SUBSEC:AGN_Fraction}). Once again, here ``ambient'' refers to the gas in the local background environment of the galaxy, whereas the ``central'' encompasses the gas within a kiloparsec of the SMBH (for further definitions and details see Section~\ref{SUBSEC:Galaxy_Properties}).

\begin{figure}
    \centering
    \includegraphics[width=0.90\linewidth]{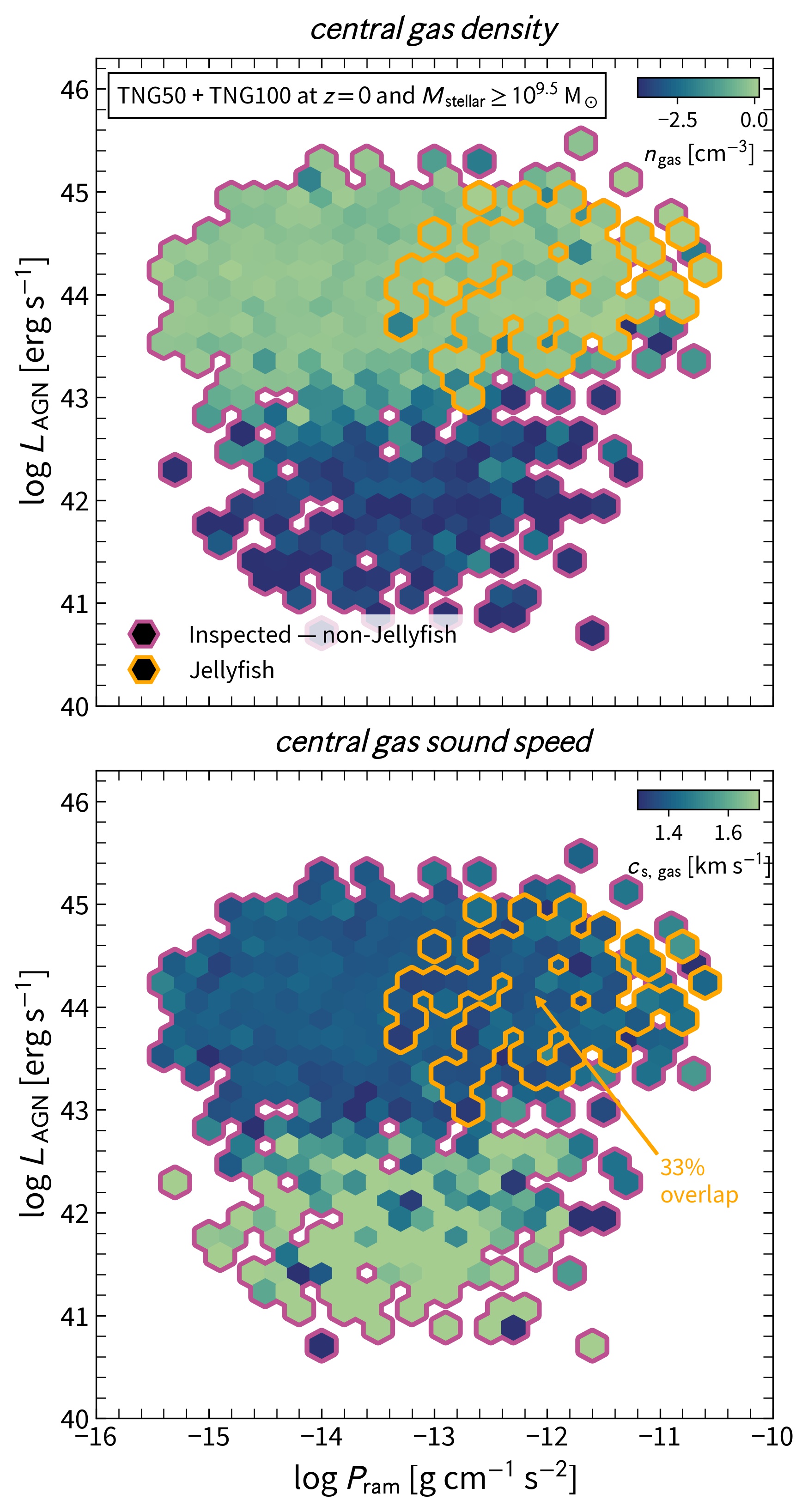}
    \caption{\textbf{\textit{Hint of ram pressure and AGN connection.}} Properties of the internal gas, within 1 ckpc of the SMBH, and the surrounding gas, in the local background environment, of jellyfish and non-jellyfish inspected satellites in TNG50 and TNG100 at $z=0$ with $M_{\rm stellar} \geq 10^{9.5} \: \rm{M}_\odot$. We show the AGN luminosity as a function of the ram pressure and indicate the average sound speed $c_s$ of the internal gas. The jellyfish (yellow outline) experience more ram pressure, have lower gas sound speeds, and higher AGN luminosities than most of the inspected satellites (pink outline; 2-D median histogram). That the jellyfish reside in only the regions associated with $P_{\rm ram}$ and $L_{\rm AGN}$ would suggest a connection between environmental stripping and SMBH activity.} 
    \label{FIG:Internal_Gas}
\end{figure}

Jellyfish galaxies as a population (above $M_{\rm stellar} \geq 10^{9.5}$ as well as $10^{10} \: \rm{M}_\odot$) reside on the end of the distribution associated with higher ram pressure, lower central gas sound speed, and thus higher AGN luminosity: this not only entails higher ambient gas densities $\rho_{\rm gas, \: LBE}$, but also higher central gas masses $M_{\rm gas}$, densities $n_{\rm gas}$, and pressures $P_{\rm gas}$ compared to the inspected satellite sample. While the distributions of the central gas properties are similar between the centrals and inspected satellites, jellyfish galaxies are distinct in that they tend toward the one mode of the gas distributions. We perform two-sample Kolmogorov-Smirnov tests and find that the jellyfish are statistically distinct from the central galaxies (purple $p$-value) and other non-jellyfish satellites with similar gas fractions (pink $p$-value in Figure~\ref{FIG:Gas_Dists}) in terms of the gas near their SMBHs, when considering the entire sample as well as well the high stellar mass bin. However, jellyfish galaxies below $M_{\rm stellar} \leq 10^{10}$ are not statistically different from inspected satellites in the low mass range, indicating that our results are largely driven by the galaxies at high stellar masses.

On the one hand, it is perhaps striking that the central regions of jellyfish have lower sound speeds, since heating from AGN feedback would in theory lead to higher temperatures and thus higher sounds speeds. However, nearly all jellyfish galaxies in our sample are in the thermal mode of AGN feedback. Although the thermal mode feedback may a priori influence the gas in the central regions to some extent, its impact is probably not significant compared to the effect of ram pressure. The lesser influence of thermal feedback may be also due to its numerical implementation, coupled with the modeling of the star-forming gas in TNG via an effective equation of state of the star-forming gas \citep{Springel_2003}. Cooling could also play some role: though we do not show the exact results here, jellyfish tend to have low cooling times below $\lesssim 0.3$ Myr, whereas the cooling times of centrals and other satellites can be up to tens or hundreds of Myr. Returning to the point raised at the beginning of Section~\ref{SEC:RP_AGN}, the fact that jellyfish galaxies as a population have high central gas densities and lower sound speeds (third column of Figure~\ref{FIG:Gas_Dists}) is actually in line with their higher AGN luminosities: according to Equation~\ref{eq:1}, the SMBH accretion rate scales with $n_\mathrm{gas}$ and inversely to $c_\mathrm{s}^{3}$. The enhanced values of $n_\mathrm{gas}$ and lower $c_\mathrm{s}$ further supports the possibility that ram pressure drives more gas from the outer regions of the galaxy toward the SMBH at its center, as we speculated in Section~\ref{SUBSEC:Mechanisms}. In the following section, we attempt to close the loop between the environment-driven effects of ram pressure and the internal activity of SMBHs in TNG by connecting properties of both ambient and central gas to AGN luminosity.

\subsection{Linking ram pressure and AGN, across satellite populations}

Figure~\ref{FIG:Internal_Gas} displays the AGN luminosity versus the ram pressure, with a 2-D median histogram: the hexagonal pixels related to inspected satellites that are {\it not} jellyfish are contoured in magenta, whereas those of jellyfish in yellow. The data is coloured according to the median central gas density (top) and the median sound speed of the central gas (bottom). Jellyfish galaxies tend to only reside in the region of the parameter space associated with high ram pressure and high AGN luminosities, and are therefore located in the upper-right region of the $L_{\rm AGN}-P_{\rm ram}$ diagram. This is in line with the fact that they are not only ram-pressure stripped, but also have high SMBH accretion rates. The jellyfish tend toward low values of $c_s$ and high $n_{\rm gas}$ relative to the rest of the sample of inspected satellites. At the population level, jellyfish are distributed toward extreme levels of ram pressure, central gas densities, and SMBH luminosity, whereas non-jellyfish do not exhibit such a preference.

However, the connection between environmental effects and SMBH luminosity in jellyfish galaxies does not fully capture how they are distinct from other gaseous satellites. Even with more ram pressure and higher AGN luminosities, jellyfish overlap with a subsample of the inspected satellites. In fact, around 33 per cent of inspected satellites have similar properties to the jellyfish in the $L_{\rm AGN}-P_{\rm ram}$ diagram: this percentage is close to the expected level of contamination from the ``Cosmological Jellyfish'' visual identification method. These inspected satellites might indeed be jellyfish, but were not classified as such because their random orientation in the images are not necessarily the optimal viewing angle for capturing ram-pressure stripped tails \citep{yun_jellyfish_2019, zinger_jellyfish_2023}.

Conversely, and without including the gaseous satellites that might have been misclassified in our sample, there are still inspected satellite galaxies in Figure~\ref{FIG:Internal_Gas} that either experience high ram pressure but do not exhibit correspondingly high AGN luminosities, or have high AGN luminosities without undergoing high ram pressure. It is apparent that ram pressure is not necessarily required to drive the SMBH accretion of all satellites, as in fact we know that other processes, especially internal ones, determines the luminosity of AGNs. We discuss the interplay of ram pressure and secular processes, such as the very feedback from SMBHs, next.

%% file: sections/sec4_discussion.tex
\section{Discussion, and on AGN Feedback in Jellyfish}                          \label{SEC:Discussion}

\begin{figure*}
    \centering
    \includegraphics[width=0.80\linewidth]{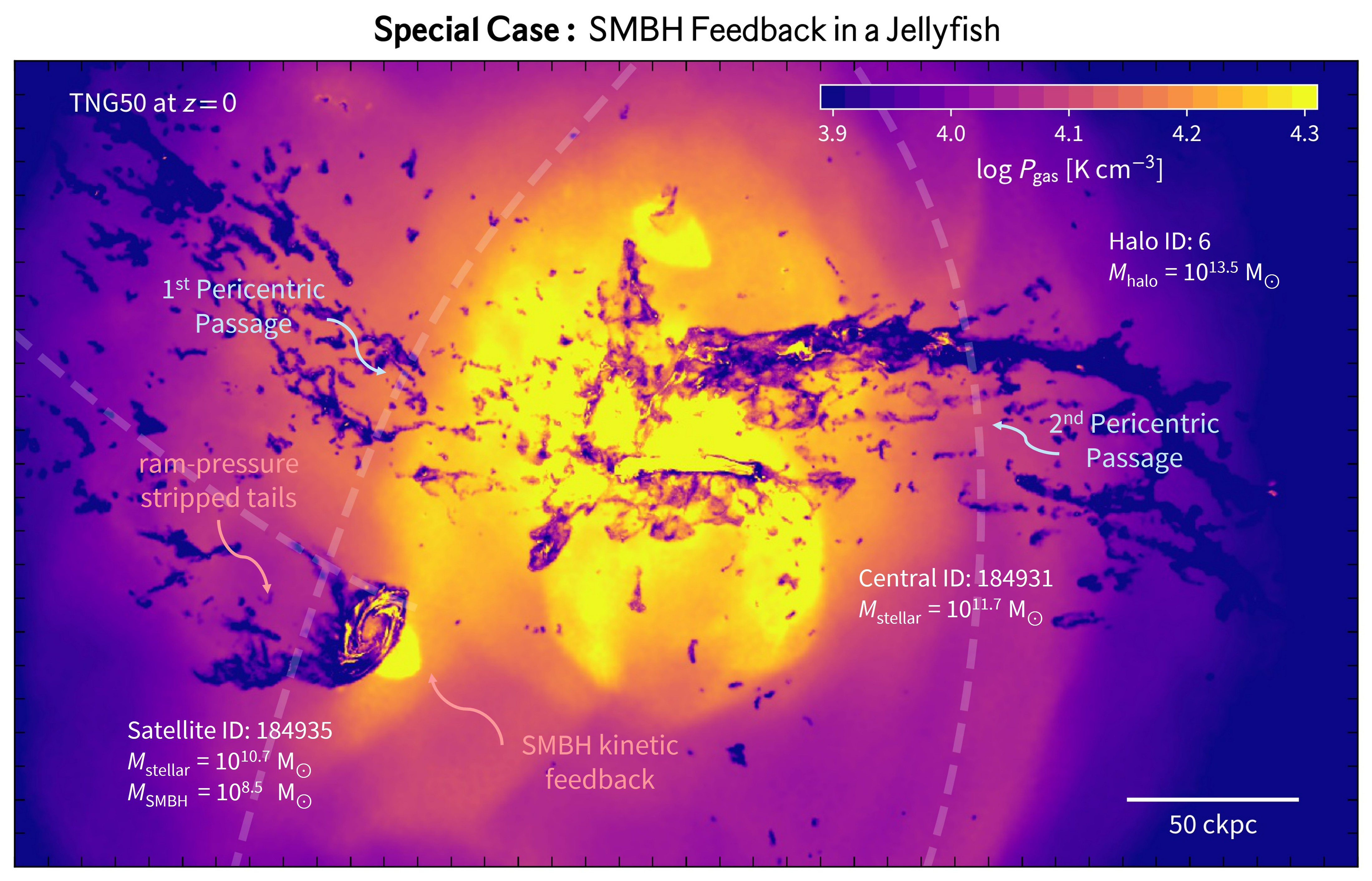}
    \caption{\textbf{\textit{The special case of a satellite with ram-pressure stripped tails and an AGN-driven bubble.}} The image shows the 2-D gas pressure projection of a satellite galaxy from TNG50 at $z = 0$, with \texttt{Subhalo ID = 184935, Snapshot 99} and a stellar mass of $M_{\rm stellar} = 10^{10.7} \: \rm{M}_\odot$. We show the trajectory of the satellite (dash line) as well as when it goes through a pericentric passage. It is orbiting a central galaxy with \texttt{Subfind ID = 184931} and a stellar mass of $M_{\rm stellar} = 10^{11.7} \: \rm{M}_\odot$, and both are residents of a larger halo structure with \texttt{Group Nr. = 6} and a host mass of $M_{\rm halo} = 10^{13.5} \: \rm{M}_\odot$. Not only does this satellite exhibit ram-pressure stripped tails, but also a bubble of high pressure gas at its leading edge. This looks similar to the eROSITA-like X-ray emitting bubbles of TNG50 Milky Way- and Andromeda-like galaxies driven by their SMBH kinetic feedback \citep[see][]{Pillepich_2021}. Interestingly, here the galaxy is a satellite and ram-pressure stripping, star formation, and gas bubbles driven by kinetic feedback from a low accretion SMBH all coexist at once.}
    \label{FIG:Special_Case}
\end{figure*}

Based on the outcome of the TNG cosmological simulations, in this paper we have provided arguments in favour of the observational hypothesis that ram pressure is a possible feeding mechanism of SMBH accretion in satellite galaxies (Section~\ref{SEC:RP_AGN}). Indeed, one explanation for the high AGN luminosities of jellyfish galaxies compared to centrals and other satellites is that the higher density and ram pressure of their cosmological environment drives or compresses more gas toward the galactic nuclei, increasing the central gas densities and decreasing their sound speeds. In turn, such enhanced physical properties of the gas in the surroundings of the SMBHs drive higher accretion rates and hence higher bolometric luminosities, helping to explain the higher AGN fractions of the jellyfish population (Section~\ref{SEC:Results}). 

On the one hand, the qualitative similarity of the outcome from TNG with the GASP observations is encouraging and is especially interesting given the fundamental simplifications of the TNG galaxy-formation model and of its SMBH physics (Section~\ref{SEC:Methodology}). In the simulations, SMBHs are modelled as sink particles and their mass accretion is modelled following a Bondi-like formulation applied to spatial scales (the often mentioned ``SMBH surroundings'') that are orders of magnitude larger than the real SMBH accretion-physics scales. Still, in the framework of these limitations, TNG shows that ram pressure, which directly affects the ambient gas around a galaxy, can also lead to effects on the gas that is more centrally concentrated within a galaxy body, i.e. the gas within a kiloparsec of the galactic nucleus that is supposed to be more closely related to the SMBH.

On the other hand, our analysis, which is based on thousands of galaxies that undergo a large diversity of evolutionary pathways, shows that there is no one-to-one, simple, or all-encompassing connection between cosmological environments and AGN activity. Firstly, whereas the large majority of TNG jellyfish galaxies exhibit the enunciated connection between ram pressure and AGN luminosity, this is not the case for all jellyfish galaxies. Moreover, we find that a fraction of satellite galaxies---among those that have been inspected, that do not necessarily exhibit tails but that still have some gas at $z=0$---may either experience high levels of ram pressure without exhibiting as high AGN luminosities as jellyfish or have similarly luminous AGNs without undergoing high ram pressure.

First and foremost, this diversity of outcomes and relationships may explain why not all observations agree as to the enhanced AGN fractions of jellyfish galaxies when compared to differing control or reference galaxy samples (Section~\ref{SEC:Introduction}). Secondly and importantly, such diversity hints at a fundamental but underrated point: the evolution of galaxies, including satellite ones undergoing environmental processes, {\it also} depends on secular and internal phenomena: determining which of those dominate across different regimes remains a central quest for studies of galaxy evolution in groups and clusters. More specifically and as anticipated at the beginning of Section~\ref{SEC:RP_AGN}, in TNG and arguably also in reality, the luminosity of AGN also depends on the feedback processes stemming from within galaxies, that is both from the SMBHs themselves as well as from star formation. In the following, we provide additional considerations on the effects of AGN feedback, and consolidate what has been already mentioned.

By removing or diluting gas in the central regions of galaxies, both feedback from star formation and feedback from AGN have been shown to modulate or even suppress the accretion rates of SMBHs in central galaxies across cosmic epochs, with stronger stellar or SMBH feedback returning ultimately smaller SMBH masses for any given halo or galaxy mass, especially towards the low-mass end and at high redshifts \citep[][in TNG]{Habouzit_2019, Truong_2021a}. In the TNG galaxy formation model, the accretion rate determines what type of feedback is exerted by SMBHs. Energy injections in the kinetic mode are activated at lower accretion rates and are more effective than thermal energy injections, which are activated at higher accretion rates, at affecting the gas in and around galaxies, chiefly by ejecting it \citep[][]{Nelson_2018, Nelson_2019, Truong_2020, Pillepich_2021}. In fact, within the TNG model, it is the kinetic SMBH feedback at low AGN luminosities that quenches star formation in massive central galaxies \citep[][]{Weinberger_2018, Terrazas_2020, zinger_ejective_2020}, marking a shift in the perspective commonly adopted until recently about quenching and AGN feedback \citep[for example,][]{Ward_2022, Bluck_2023}. 

Notwithstanding details related to the exact choices and implementations of any galaxy formation models, it is clear that the growth, and hence accretion rates at any given time, of SMBHs is determined by a host of processes. In the case of central TNG galaxies, low-mass ($M_{\rm stellar} \lesssim 10^{8.0-8.5} \: \rm{M}_\odot$) or high-redshift ($z \gtrsim2$) SMBHs predominantly grow through accretion in the thermal mode, whereas more massive SMBHs have a rapid accretion phase at high redshifts until a certain mass and then build up most of their mass by way of mergers at later times, with kinetic feedback effectively self-limiting their growth via gas accretion again \citep{Weinberger_2018}. The picture for satellite galaxies can only be more complex, given that additional processes determine the availability of gas for SMBH accretion, chiefly stripping from tides and ram pressure. Additionally, feedback-driven outflows may further enhance the stripping of gas that would otherwise be available for accretion onto the SMBH \citep[see also][as examples]{ricarte_link_2020, rohr_jellyfish_2023}. 

In Appendix~\ref{A:AGN_FB}, we provide an overview of the accretion rate regimes of $z=0$ satellite and jellyfish galaxies across time, noting that galaxies may transition from the thermal to kinetic mode of SMBH feedback before or after infall, also depending on their mass and on pre-processing \citep[e.g.][]{donnari_quenched_2021}. Importantly, we find that the largest majority of jellyfish galaxies, those towards the low-mass end, have been in thermal mode for their entire life, even before the present-day as per Section~\ref{SEC:Results}. This suggests that, in addition to a physical causation between ram pressure and enhanced AGN luminosity, jellyfish galaxies are necessarily biased towards accretion and feedback in the thermal mode to even have the chance to retain gas and to appear ram-pressure stripped. In contrast, the majority of satellite galaxies (especially towards the low-mass end of the considered sample), are devoid of gas and have very low SMBH accretion rates as well as luminosities, chiefly because of environmental effects such as ram-pressure stripping, rather than feedback, forcing their SMBHs to transition to the kinetic mode despite having SMBH masses much smaller than the typical central galaxy in kinetic mode.

With this picture in mind, we speculate that those inspected satellite galaxies with low AGN luminosities despite high ram pressure are higher-mass galaxies in the kinetic mode that have transitioned to low accretion rates because of secular feedback rather than environmental effects. The second scenario of inspected satellites, with low levels of ram pressure but nevertheless high AGN luminosities, is less straightforward to explain and serves as a reminder that ram pressure is not required for elevated SMBH accretion in satellites.


\subsection{The special case of SMBH kinetic feedback in a jellyfish}

At the high-mass end of satellite galaxies, we can expect SMBH-driven outflows to be a present phenomena consistent with observations. \citet{ricarte_link_2020} show simulations where ram pressure enhances SMBH accretion and triggers AGN feedback in the form of heating and outflows; they speculate that both the SMBH and environment thus quench star formation. From the perspective of the TNG simulations, \citet{donnari_quenched_2021} has demonstrated that the kinetic mode of AGN feedback is chiefly responsible for quenching star formation, not only in massive central galaxies, but also in massive satellites that reside in groups and clusters. At the same time, these galaxies may still exhibit signatures of environmental stripping \citep{joshi_fate_2020}. We close this discussion by showcasing a suggestive example of a TNG galaxy exhibiting both ram-pressure stripped tails as well as a SMBH-driven bubble.

\begin{figure}
    \centering
    \hspace{1em}
    \ContinuedFloat
    \includegraphics[width=0.92\linewidth]{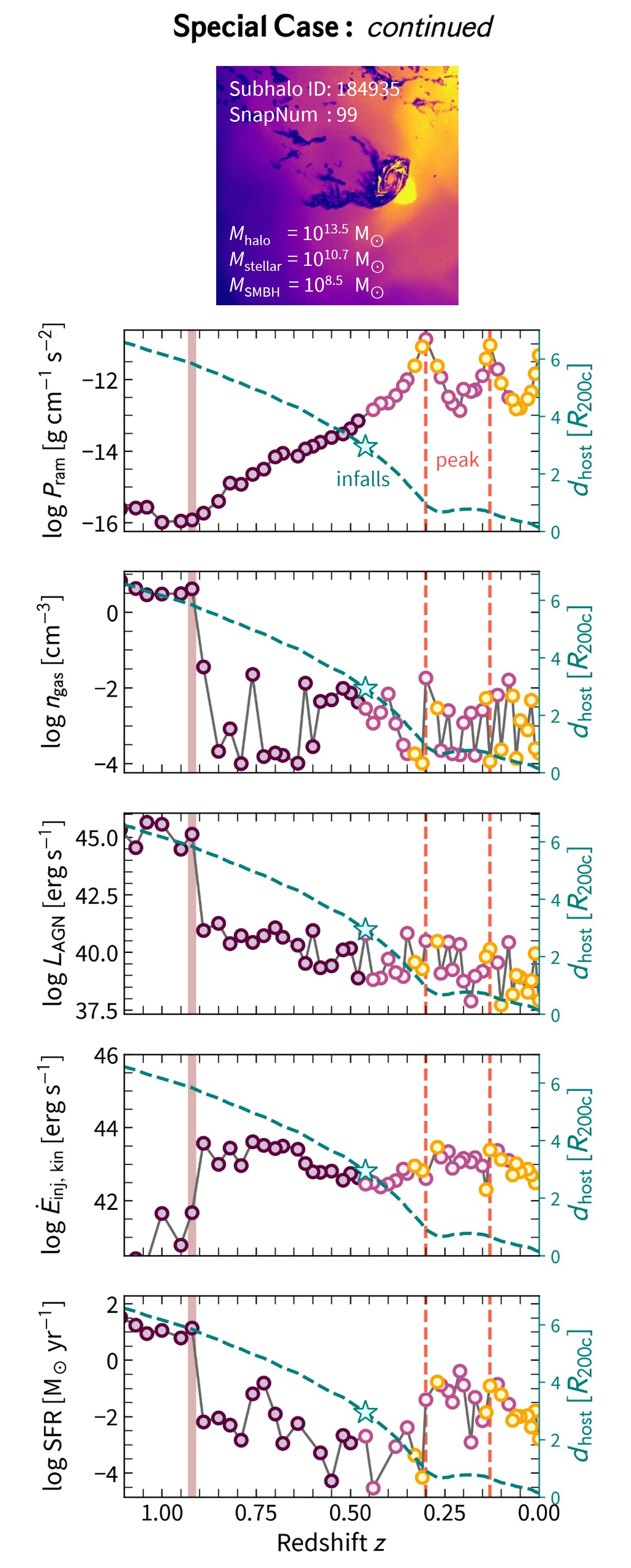}
    \contcaption{\textbf{\textit{satellite with ram-pressure stripped tails and AGN feedback.}} The redshift evolutions of the satellite properties are shown in this order: ram pressure, central gas density, AGN luminosity, AGN kinetic energy injected, and star formation rate; we show the host-centric distance on a secondary axis (dashed green line). We also track whether the galaxy is a jellyfish (yellow circle), inspected satellite (pink circle), or central (purple circle) at each snapshot. We mark when the galaxy's SMBH transitions from the thermal to kinetic feedback mode (dark red line), infalls (green star), and experiences peaks in ram pressure (dashed red lines). After transitioning to the kinetic mode, the galaxy loses a large fraction of its internal gas reservoir. The episodes of feedback coincide with bursts in the SFR and AGN luminosity. }
    \label{FIG:Special_Case_Continued}
\end{figure}

Figure~\ref{FIG:Special_Case} shows a spectacular manifestation of AGN feedback in a ram-pressure stripped galaxy with $M_{\rm stellar} = 10^{10.7} \: \rm{M}_\odot$ from TNG50 at $z=0$, in a 2-D projection of gas pressure projection. Located at the lower left corner of the image, the galaxy is a satellite of a massive central galaxy with $M_{\rm stellar} = 10^{11.7} \: \rm{M}_\odot$, in a host halo with $M_{\rm host} = 10^{13.5} \: \rm{M}_\odot$. It leaves behind trails of dark, low-pressure gas, much like the tails of a jellyfish galaxy. As a matter of fact, this satellite was not classified as a jellyfish due the random-viewing angle of the ``Cosmological Jellyfish'' project---for the case of this galaxy alone, we classify it as a jellyfish based on its edge-on orientation. More striking is the bubble of high-pressure gas emitting from the front and center of the jellyfish face. The galaxy has a SMBH with mass $M_{\rm SMBH} = 10^{8.5} \: \rm{M}_\odot$ and an accretion $L_{\rm AGN} = 10^{37.9} \: \rm{erg \: s}^{-1}$ luminosity, placing it in the SMBH kinetic feedback regime. Given the similarity of the pressurized bubble of gas above the disk of this galaxy with those emerging in the majority of Milky Way and Andromeda-like galaxies in TNG50 at $z=0$ \citep{Pillepich_2021} as well as to the eROSITA bubbles of our own Galaxy \citep{Predehl_2020}, we argue that the bubble in the satellite galaxy of Figure~\ref{FIG:Special_Case}  is due to the kinetic energy injections from its SMBH. 

The galaxy of Figure~\ref{FIG:Special_Case} is special, and a counter example to the bulk of the galaxies revealed in this paper, for three reasons. First, it is a massive jellyfish---in fact, a Milky Way-mass satellite---noting that jellyfish become rarer toward larger masses (see Figure~\ref{FIG:SMBH_Dists} and Appendix~\ref{A:AGN_FB}). Second, it is a jellyfish despite hosting a SMBH that accretes at low rates and can hence impart mechanical feedback---such feedback is typically efficient at expelling gas from galaxies, ultimately reducing their chances of even being inspected for tails. Third, the SMBH kinetic feedback manifests as an eROSITA-like bubble despite the galaxy orbiting through the dense intragroup medium of its host and in the direction of the bubble's extent.

To better understand this complexity, we examine the redshift evolution of the satellite's properties on the continuation of Figure~\ref{FIG:Special_Case}, in the following order: the ram pressure $P_{\rm ram}$, central gas density $n_{\rm gas}$, AGN luminosity $L_{\rm AGN}$, energy injected as SMBH kinetic feedback $\dot{E}_{\rm inj, \: kin}$, and star formation rate SFR. Like in Figure~\ref{FIG:RP-AGN}, we track the galaxy's infall into the host halo through measuring its host-centric distance, and the time at which it becomes a satellite. Now, we also indicate the transition from thermal to kinetic feedback and whether it is a jellyfish or not based on a re-inspection of the satellite-halo system at a more optimal viewing angle, shown in the image. The classification of the satellite as a jellyfish is still inconsistent between snapshots towards $z = 0$, likely due to intense effects of SMBH feedback and tidal forces that could disrupt the tails.

At $z=1$, well before infall, the galaxy is a central (purple markers), relatively isolated from dense environments and ram-pressure effects. With a large central gas supply, it is star-forming and its SMBH is actively accreting in the thermal mode. At this initial stage, the galaxy's SMBH has an exceptionally high AGN luminosity of $L_{\rm AGN} \simeq 10^{45} \: \rm{erg \: s^{-1}}$. After building enormous amounts of mass, the SMBH enters the kinetic feedback mode at $z \simeq 0.92$. The immediate aftermath of this transition is a burst of outflows. There is a surge of kinetic energy injections until $z \sim 0.5$: the galaxy loses most of its star-forming and SMBH fuel as its central gas density decreases, and thus its star formation starts to quench and the AGN luminosity declines to its present-day levels. The galaxy becomes a satellite (and is inspected for jellyfish-like tails, pink markers) at $z \simeq 0.45$. As it enters its host halo, the satellite experiences increased ram-pressure, which peaks for the first time at $z \simeq 0.3$. It orbits past the central galaxy before falling back at $z \simeq 0.13$, where it experiences a second peak in ram pressure. Lastly at $z \simeq 0$, it approaches yet another pericentric passage, for a last peak in ram pressure. Through the second peak, jellyfish-like tails have become stable (or rather, visually identified consistently through time amid projection effects; yellow markers). After infall, and during the rises and falls of ram pressure, the central gas density and AGN luminosity oscillate rapidly. The kinetic energy injected swells, and the star formation is rejuvenated.

This special case transitions from the thermal to kinetic mode first, then falls into its $z=0$ host, and finally becomes a jellyfish last. In fact, galaxies along a similar path switch from the thermal mode of AGN feedback while they are still centrals. The extended exposure to kinetic energy injections is destined to lessen the gravitational binding energy of their outflowing gas. By the time they enter their host halos and become satellites, it is easier for ram pressure to strip the outer layers of gas to form jellyfish-like tails. During this time, ram pressure can also compress gas in the inner regions, feeding SMBH accretion (like in Section~\ref{SEC:RP_AGN}) as well as star formation. Simultaneously, the kinetic feedback combined with stellar feedback is expected to drive outflows and keep the SMBH accretion low as a competing effect. The galaxies appear as jellyfish before $z=0$ as a likely result of both AGN feedback and ram-pressure, but do not necessarily remain so. Eventually, the environment and the SMBH feedback injected after infall could remove too much of the gas for the satellites to be considered as jellyfish by the present day.

%% file: sections/sec5_conclusion.tex

\section{Conclusions}                        \label{SEC:Conclusions}

In this work, we utilize the IllustrisTNG cosmological simulations (specifically, TNG50 and TNG100) to investigate SMBHs in jellyfish galaxies, as identified by the ``Cosmological Jellyfish'' citizen-science project, with stellar masses of $M_{\rm{stellar}} \geq 10^{9.5} \: \rm{M}_\odot$ at redshift $z = 0$. We compare the jellyfish to other galaxies, namely centrals, satellites in general, and inspected satellites that are not necessarily ram-pressure stripped (see Figure~\ref{FIG:Sample_Selection}). Our findings support a connection between the environment and SMBH activity, consistent with some observational evidence, and are outlined here.

\begin{enumerate}
    \item \textbf{\textit{According to the TNG simulations, jellyfish galaxies have more active SMBHs than centrals and other satellites of the same mass}} (Figure~\ref{FIG:SMBH_Dists}). The SMBHs of jellyfish have overall higher accretion rates and luminosities above $L_{\rm AGN} \gtrsim 10^{42} \: \rm{erg \: s}^{-1}$ compared to galaxies that are not affected by the environment. They tend to have lower stellar masses $M_{\rm{stellar}} \lesssim 10^{10.8} \: \rm{M}_\odot$ and SMBH masses at $M_{\rm{SMBH}} \lesssim 10^{8} \: \rm{M}_\odot$, and are in the thermal mode of AGN feedback almost without exception (see Section~\ref{SEC:Discussion} for more details on the AGN feedback modes of jellyfish), whereas this is only the case for 85 per cent of centrals and 65 per cent of satellites. \\

    \item \textbf{\textit{The AGN fraction in jellyfish is similar to or higher than in other gaseous satellites}} (Figure~\ref{FIG:AGN_Fraction}). Between $50 - 70$ per cent of jellyfish host an AGN with $L_{\rm AGN} \geq 10^{44} \: \rm{erg \: s}^{-1}$, which increases with larger stellar masses. This qualitatively agrees with the observations of the GASP team \citep{poggianti_ram-pressure_2017, peluso_exploring_2022}, and is especially true at high stellar masses, $M_{\rm{stellar}} \gtrsim 10^{10} \: \rm{M}_\odot$. Above this range, the AGN fraction declines for other samples while the jellyfish are even more likely to host an active SMBH compared to centrals and other gaseous satellites. \\

    \item \textbf{\textit{The ram pressure exerted by the dense cosmological environments can increase the density of gas} near the jellyfish nuclei, lowering their sound speeds, and enhancing the SMBH accretion and hence its luminosity} (Figure~\ref{FIG:Internal_Gas}). We find that when 65 per cent of jellyfish galaxies fall in toward their halo centres, they not only experience a peak in ram-pressure, but show increases in central gas densities and AGN luminosity near the time of or slightly after their first pericentric passage (Figures~\ref{FIG:RP-AGN} and~\ref{FIG:RP-AGN_Cumulative}). We tentatively show that ram pressure feeds SMBH accretion through compressing and driving inflows of gas toward the galaxy center (Figure~\ref{FIG:Mechanisms}). Indeed, compared to other gaseous satellites and central galaxies, and in terms of the gas near the SMBHs, jellyfish are significantly biased towards higher gas masses, pressures, and densities
    (Figure~\ref{FIG:Gas_Dists}).
\end{enumerate}

Our analysis of thousands of TNG50 and TNG100 galaxies from the IllustrisTNG project therefore seems to support the observationally-proposed idea that ram pressure enhances SMBH accretion in jellyfish through influencing the internal gas of the galaxies. However, this is not always the case because the secular evolution of satellites, even prior to becoming such, impact their AGN luminosity at any given time of inspection. Indeed, the evolutionary paths are diverse and internal and environmental processes are intertwined with varying impacts, so that one track does not fit all. This is epitomized by the special case of a Milky Way-mass jellyfish galaxy from TNG50 exhibiting an eROSITA-like bubble \citep{Pillepich_2021} driven by kinetic-feedback outflows (Figure~\ref{FIG:Special_Case}). TNG model aside, it is plausible that for jellyfish to even have gas available for stripping as well as for accretion, their SMBH feedback must not be too ejective, allowing them to host luminous AGN. In addition, jellyfish are more likely to host a luminous AGN than satellites of the same mass largely due to the fact that the majority of satellite galaxies are gas-poor and thus have lower SMBH accretion rates. 

The implications of AGN feedback for star formation in jellyfish are even more complex and somewhat unexpected. The quenching of star formation due to SMBH kinetic feedback in massive satellites \citep{donnari_quenched_2021} does not preclude spectacular ram-pressure stripped tails (Figure~\ref{FIG:Special_Case}): it is plausible that the removal of gas via ram-pressure stripping can enable the transition to the low-accretion, kinetic mode. Similarly, the elevated AGN fractions observed in jellyfish galaxies do not contradict the broader trend that gas accretion onto SMBHs in group and cluster satellites is typically suppressed compared to their field counterparts \citep{joshi_fate_2020}. In a similar fashion, bursts of star formation in jellyfish galaxies \citep{goller_jellyfish_2023}---roughly coinciding with episodes of AGN luminosity---can still occur even as the satellite population as a whole undergoes quenching. Eventually, jellyfish galaxies are likely to lose much of their gas due to a combination of SMBH feedback and ram-pressure stripping. This will ultimately suppress both their star formation rates and AGN activity, as is seen in satellites more generally.

The results presented in this paper, especially those more closely connected with the thermal and kinetic mode of SMBH feedback, the implementation of the Bondi accretion rate, as well as the positioning and seeding of black holes, are necessarily TNG model-dependent, and could change with more physically-motivated, higher-resolution simulations. It is important to model SMBH growth and feedback in detail when analysing the evolution of satellites and jellyfish in future theoretical experiments. Additionally, the ambient gas density cannot be directly inferred from measurements of the ram pressure stripping without accounting for feedback from star formation and AGN. It would be insightful to explore how the connection between ram pressure and SMBHs varies across galaxy formation models too.

This work further corroborates the importance of internal processes in massive satellite galaxies that are also strongly influenced by their environment, as we find evidence that SMBH activity and ram-pressure stripping could work in tandem. It is therefore crucial to consider both the environment and SMBHs for a more complete picture of satellite galaxy evolution. Similar mechanisms may potentially have operated in the early Universe, where dense conditions fostered more frequent interactions. Shedding light on jellyfish galaxies allows us to continue to learn more about the co-evolution of galaxies, SMBHs, and their environments.

%% file: sections/sec6_appendix.tex

\section{The ``Missing'' Black Holes in TNG}
                                     \label{A:Missing_BHs}

Section \ref{SUBSUBSEC:SMBH} mentions that some of the galaxies in TNG50 and TNG100 with stellar masses exceeding $10^{8.3}$ and $10^{9.5} \: \rm{M}_\odot$, respectively, do not contain an SMBH. This may be due to two possible reasons related respectively to the SMBH seeding and positioning model of TNG: namely, they may have never gotten a SMBH because they were not massive enough or they have lost their SMBH.

\begin{figure}
    \centering
    \includegraphics[width=0.95\linewidth]{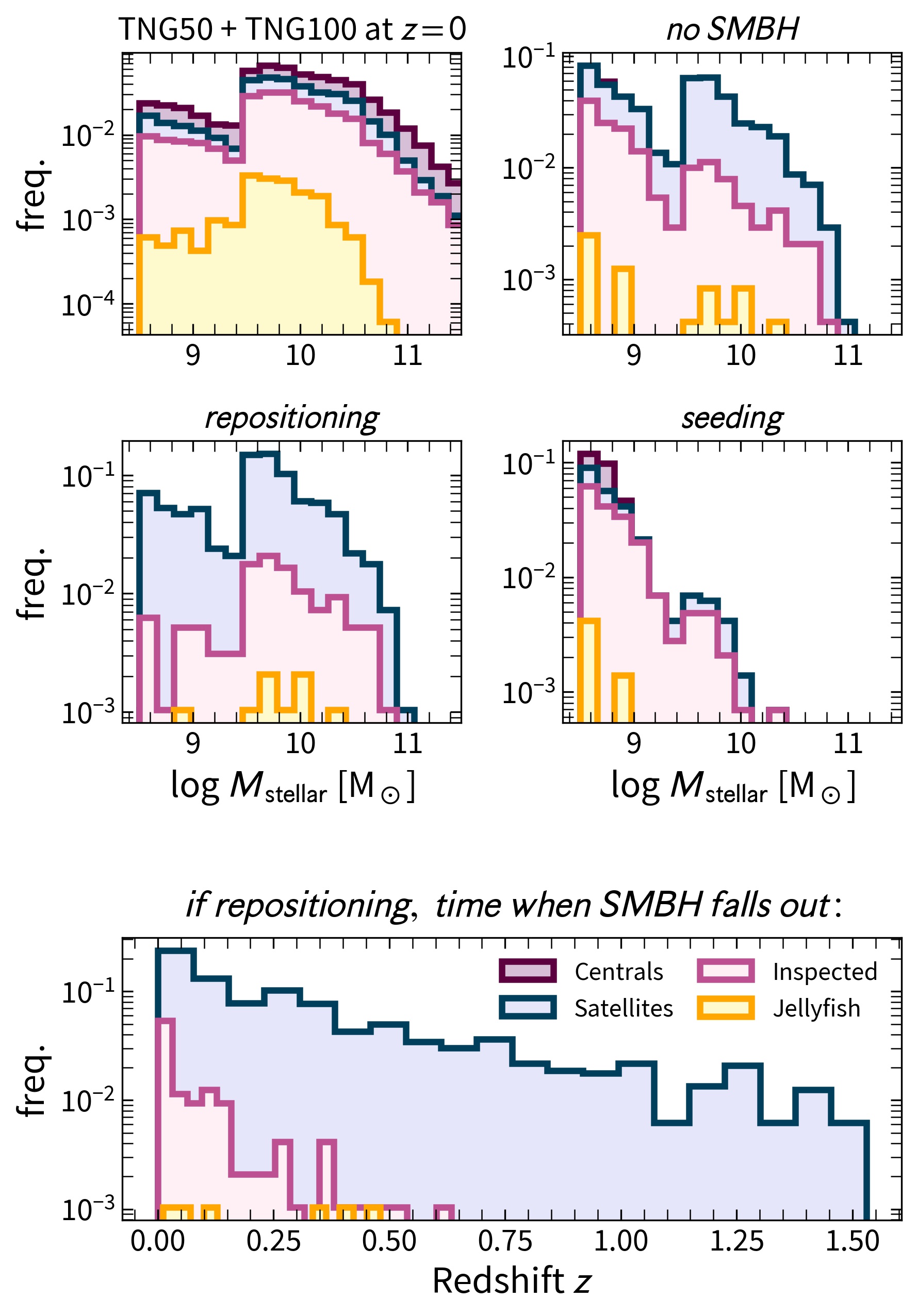}
    \caption{\textbf{\textit{Fate of SMBHs in IllustrisTNG --- ``Missing'' Black Holes.}} The top panels display the frequency of centrals (purple), satellites (blue), inspected satellites (pink), and jellyfish (yellow) from TNG50 and TNG100 among all galaxies whose SMBHs experience one of two fates, in stellar mass bins. For reference, we show the total number of galaxies across the subsamples of this paper (top left) and those that do not contain an SMBH at $z = 0$ (top right). The galaxies that do not have an SMBH either lost it at some point in their lifetimes (middle left) or they never seeded a black hole particle from the start (middle right). For the galaxies who lost their SMBH in the past, we show the distribution of the redshift when the black hole particle has ``fallen out'' of the subhalo (bottom). This typically happens after $z \sim 2.5$, probably due to a close encounter with a nearby massive neighbour. Most galaxies that never seed an SMBH have $M_{\rm stellar} \leq 10^{9.5} \: \rm{M}_\odot$ in TNG50. See also Table~\ref{TAB:Missing_BHs} for statistics on the fate of missing SMBHs.}
    \label{FIG:Missing_BHs}
\end{figure}

The simulations place a black hole particle with a seed mass of $M_\mathrm{seed} = 8 \times 10^5 \: h^{-1} \: \mathrm{M}_\odot \simeq 1.2 \times 10^{6} \: \mathrm{M}_\odot$ in a friends-of-friends (FoF) halo when it exceeds $M_\mathrm{FoF} = 5 \times 10^{10} \: h^{-1} \: \mathrm{M}_\odot  \simeq 7.4 \times 10^{10} \: \mathrm{M}_\odot$, at the center of its gravitational potential well. In one possible fate, the lowest-mass galaxies below $M_{\rm stellar} \leq 10^{9.5} \: \rm{M}_\odot$ never achieve this criteria and hence never seed an SMBH. In the other possible fate, some galaxies have lost their SMBH: if a black hole particle is seeded in a low-mass halo that is then stripped or disrupted, it can be repositioned to to the potential center of a new halo. This can happen during close encounters with nearby, more massive galaxies. The percentage of galaxies that do not contain a SMBH due to either of these reasons is reported in Table~\ref{TAB:Missing_BHs} and Figure~\ref{FIG:Missing_BHs}, which also shows when SMBHs typically ``fall out'' due to 
repositioning.

The lack of SMBHs in these galaxies is an artificial result of the simulation. To minimize the number of objects in our sample that have never seeded a black hole, we only consider galaxies with stellar masses above the threshold $M_{\rm stellar} \geq 10^{9.5} \: \rm{M}_\odot$ in our final selection. For the remaining galaxies that still have no SMBH due to repositioning, we assign a ``fudge'' black hole with a mass and accretion rate randomly selected from the Gaussian distribution of SMBHs in the sample at $z = 0$. \citet{Goubert_2024} run into a similar issue, where a significant fraction of low-mass satellites lack black holes in simulations, IllustrisTNG included. They show that reassigning ``fudge'' values to the affected subhalos does not lead to different end results. In this work, we choose to de-select subhalos below the stellar mass cut and manually assign a SMBH mass and accretion rate for the remaining few galaxies without black holes, as a compromise between sample purity and completeness. However, it is important to note that as a consequence of the TNG model, we have to exclude some low-mass jellyfish galaxies.

\section{AGN Luminosities and Fractions} \label{A:L_AGN}

In Figure~\ref{FIG:AGN_Fraction} of Section~\ref{SUBSEC:AGN_Fraction}, we show the AGN fraction of galaxies in various cosmological environments in stellar mass bins. We calculate the AGN fraction as the fraction of galaxies with AGN luminosities above $L_{\rm AGN} \geq 10^{44} \: \rm{erg \: s}^{-1}$. However, for reference, Figure~\ref{FIG:AGN_Fraction_Cuts} shows the results of alternative luminosity cuts. Across all luminosity thresholds, jellyfish galaxies (yellow) consistently show the highest AGN fractions, suggesting a strong connection between jellyfish classification and AGN activity. The trend is most prominent at the highest AGN luminosities ($L_{\rm AGN} \geq 10^{44} \: \rm{erg \: s}^{-1}$).

\renewcommand{\thetable}{A\arabic{table}}
\setcounter{table}{0}
\begin{table}
\caption{\textbf{\textit{Percentage of TNG50 and TNG100 galaxies that do not contain an SMBH at $z = 0.0$}} because they have either ``lost'' it due to repositioning, or their black hole particle never seeded. Imposing a stellar mass cut of $M_{\rm stellar} \geq 10^{9.5} \: \rm{M}_\odot$ for both TNG50 and TNG100 allows us to exclude the majority of the galaxies that do not host an SMBH for these reasons. For the remaining galaxies with missing black hole particle, we assign a ``fudge'' SMBH mass and accretion rate from a Gaussian distribution.}
\begin{tabular}{lcccc}
\hline
\multicolumn{1}{l|}{\textit{Percentage \%}} & \multicolumn{2}{c|}{\textit{Lost SMBH to Neighbour}} & \multicolumn{2}{c}{\textit{Never Seeded SMBH}} \\ \hline
\multicolumn{1}{l|}{\textbf{Sample}}                 & \textbf{TNG50}                 & \multicolumn{1}{c|}{\textbf{TNG100}}  & \textbf{TNG50}               & \textbf{TNG100}               \\ \hline

\multicolumn{1}{l|}{\textit{\textbf{Galaxies}}}      & 7.32 $\rightarrow$ \textbf{3.14}          & \multicolumn{1}{c|}{\textbf{5.13}}               & 25.1 $\rightarrow$ \textbf{0.13}                      & \textbf{0.23}                            \\
\multicolumn{1}{l|}{\textit{\textbf{Centrals}}}      & 0.39 $\rightarrow$ \textbf{0.11}          & \multicolumn{1}{c|}{\textbf{0.06}}               & 23.2 $\rightarrow$ \textbf{0.11}                      & \textbf{0.00}                            \\
\multicolumn{1}{l|}{\textit{\textbf{Satellites}}}    & 18.5 $\rightarrow$ \textbf{7.99}          & \multicolumn{1}{c|}{\textbf{12.6}}               & 26.4 $\rightarrow$ \textbf{0.17}                      & \textbf{0.58}                            \\
\multicolumn{1}{l|}{\textit{\textbf{Inspected}}}     & 2.54 $\rightarrow$ \textbf{1.95}          & \multicolumn{1}{c|}{\textbf{2.91}}               & 28.9 $\rightarrow$ \textbf{0.00}                      & \textbf{0.61}                            \\
\multicolumn{1}{l|}{\textit{\textbf{Jellyfish}}}     & 3.13 $\rightarrow$ \textbf{1.69}          & \multicolumn{1}{c|}{\textbf{2.86}}               & 19.5 $\rightarrow$ \textbf{0.00}                      & \textbf{0.00}                            \\ \hline
\textbf{TNG50}           & \multicolumn{4}{l}{$M_{\rm stellar} \geq 10^{8.3}\: \rm{M}_\odot$ → $\mathbf{\mathit{M}_{\rm stellar} \geq 10^{9.5} \: \rm{M}_\odot}$}                       \\
\textbf{TNG100 }         & \multicolumn{4}{l}{$\mathbf{\mathit{M}_{\rm stellar} \geq 10^{9.5} \: \rm{M}_\odot}$}  
\\ \hline
\end{tabular}
\label{TAB:Missing_BHs}
\end{table}
\renewcommand{\thetable}{A\arabic{table}}
\setcounter{table}{0}

\renewcommand{\thefigure}{B\arabic{figure}}
\setcounter{figure}{0}
\begin{figure*}
    \centering
    \includegraphics[width=\linewidth]{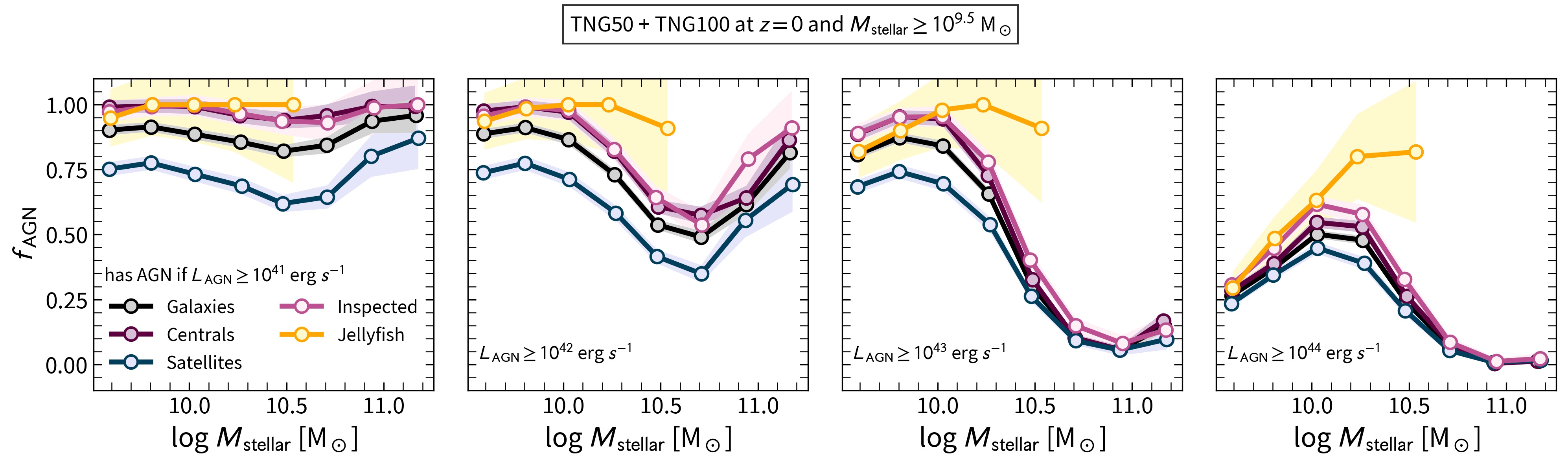}
    \caption{\textbf{\textit{AGN fractions for different AGN luminosity cuts.}} We quantify the AGN fraction in stellar mass bins for jellyfish (yellow), inspected satellites (pink), all satellites (blue), centrals (purple), and the whole populations of galaxies (black) in TNG50 and TNG100 at $z = 0$ with $M_{\rm stellar} \geq 10^{9.5} \: \rm{M}_\odot$. From left to right, we consider a galaxy to host and AGN if it has an AGN luminosity greater than $L_{\rm AGN} \geq 10^{41}, 10^{42}, 10^{43},$ and $10^{44}$ erg s$^{-1}$, our fiducual choice.}
    \label{FIG:AGN_Fraction_Cuts}
\end{figure*}
\renewcommand{\thefigure}{B\arabic{figure}}
\setcounter{figure}{0}


\section{AGN feedback modes of satellites} \label{A:AGN_FB} 

In Figure \ref{FIG:AGN_FB_Class}, we check the SMBH evolution of the jellyfish and inspected satellites, and track their AGN feedback mode as well as their jellyfish classification. In the figure we focus and report only TNG50 galaxies but the following statements apply to both TNG50 and TNG100 galaxies. The top row shows the probability that a progenitor galaxy is classified as a central, satellite, inspected satellite, or a jellyfish as a function of redshift, for three samples at the present-day: jellyfish (left column), satellites with SMBHs in the thermal feedback mode (middle column), and satellites with SMBHs in the kinetic mode (right column). In the bottom rows, we show the evolutionary tracks on the $f_{\rm Edd}-M_{\rm SMBH}$ diagram of two or three example galaxies in each of these $z=0$ samples (jellyfish, satellites in SMBH thermal feedback and satellites in SMBH kinetic feedback, from left to right), to indicate their feedback activity over time.

In the case of $z=0$ jellyfish galaxies, the most common scenario is the one in the top left track: jellyfish have been in the thermal mode for their entire lifetimes, both when centrals before becoming satellites or inspected satellites as well as satellites. There are fewer than ten (8 out of 242 $\simeq$ 3 per cent) among all the jellyfish galaxies in TNG50 and TNG100 that are in the kinetic mode at $z = 0$. To this rare category belongs the case in the bottom left panel: the galaxy infalls and then transitions to the kinetic mode. Its gas reservoir is in flux as it switches between being a gaseous (i.e. inspected) versus a gas-poor satellite, and it only becomes a jellyfish at the present day. In most cases, ram-pressure stripped galaxies ``need to be'' in the thermal mode to be inspected and to appear as jellyfish in the ``Cosmological Jellyfish'' Zooniverse classification. Because the kinetic, low-accretion mode feedback in TNG is considerably more effective than the thermal one at expelling gas from galaxies' centres \citep{zinger_ejective_2020, zinger_jellyfish_2023}, it is more probable for satellites to retain gas, to have the chance to be inspected, and to possibly be ram-pressure stripped if they are in thermal AGN feedback mode and have not yet transitioned to the kinetic mode. This is important to note because whether a galaxy may even be inspected depends not only on environmental effects but also on internal processes such as their SMBH feedback itself. 

\renewcommand{\thefigure}{C\arabic{figure}}
\setcounter{figure}{0}
\begin{figure*}
    \centering
    \includegraphics[width=0.80\linewidth]{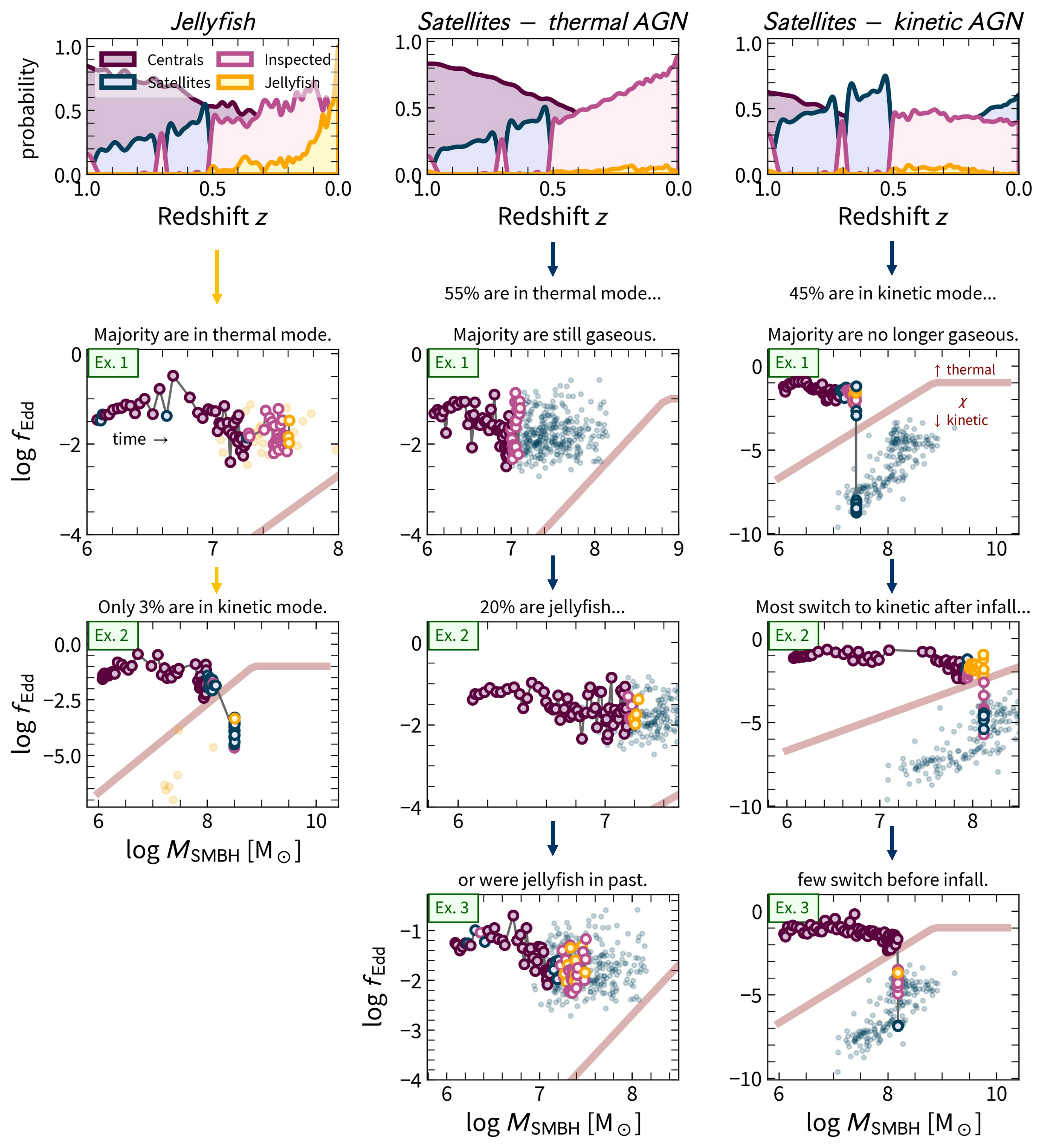}
    \caption{\textbf{\textit{The modes of SMBH feedback across cosmic times, and their various evolutionary pathways...}} \textit{(caption continued on next page)}}
    \label{FIG:AGN_FB_Class}
\end{figure*}
\renewcommand{\thefigure}{C\arabic{figure}}
\setcounter{figure}{0}

\renewcommand{\thefigure}{C\arabic{figure}}
\setcounter{figure}{1}
\begin{figure*}
    \centering
    \contcaption{... that lead to jellyfish and satellites with SMBHs in the thermal or kinetic mode of feedback. The top row shows the probability of a galaxy to be a central (purple), satellite (blue), inspected satellite (pink), or jellyfish (yellow), as a function of redshift $z$, following the main progenitor branches of three samples from TNG50 at $z = 0$: jellyfish (left), satellites with an AGN in the thermal mode (middle), and satellites with an AGN in the kinetic mode (right). Note that the ``Cosmological Jellyfish'' dataset does not include all snapshots beyond $z =0.5$, and so we do not find inspected satellites for all of these redshifts. The lower rows show the evolution of randomly-selected galaxies from each sample on the $f_{\rm Edd}-M_{\rm SMBH}$ diagram, with their classification indicated by colour at each snapshot, compared to the total sample of jellyfish or satellites (faint circles in the background). Each represents a possible evolutionary path. The thermal-to-kinetic feedback boundary is shown with a faint thick red line. In the thermal mode, the galaxies evolve from being centrals to becoming jellyfish or inspected satellites. Meanwhile, satellites in the kinetic mode may previously have been jellyfish before losing their gas to AGN feedback or the environment.}
\end{figure*}
\renewcommand{\thefigure}{C\arabic{figure}}
\setcounter{figure}{0}

Whereas jellyfish galaxies are, for the most part, in the high-accretion, thermal mode of AGN feedback, the evolutionary pathways of the population of satellite galaxies in general are more various. In fact, 55 per cent of satellites are in the thermal mode (middle column) and 45 per cent are in the kinetic mode at $z = 0$ in the TNG simulations and depending on the galaxy or SMBH mass. Satellites whose SMBHs are in the thermal mode at $z=0$ are in 80 per cent of the instances still sufficiently gaseous to be inspected, and the majority of them do not exhibit, at the time of inspection, ram-pressure stripped tails (first example). In contrast, at least 20 per cent are either currently jellyfish (second example) or were jellyfish in the past, or at least inspected to be such at the available snapshots (third example). For these satellites, ram pressure as opposed to AGN feedback is primarily responsible, not only for the formation of jellyfish tails, but also for depleting their gas. 

The largest majority of satellites whose SMBHs are in the kinetic mode at $z = 0$ (right column) are no longer gaseous due to AGN feedback ejecting their internal gas reservoirs (first example). Although most satellite galaxies in the kinetic mode are not currently gaseous enough to have ram-pressure stripped tails, they may have been jellyfish in the past. As in the case of satellites in the thermal mode, some fraction of satellites in the kinetic mode at $z = 0$ were once jellyfish---however, the majority of cases switched to the kinetic mode \textit{after} infalling into their host halo (second example). Indeed, they probably transitioned to the SMBH kinetic mode feedback because ram pressure stripped much of their gas, hampering accretion onto their SMBHs and hence moving them below the thermal-to-kinetic mode threshold. Meanwhile, up to 10 per cent of $z = 0$ satellites in the kinetic mode that used to be jellyfish switched the kinetic mode \textit{before} infall like in the very special case of SMBH feedback we show in Figure~\ref{FIG:Special_Case} (third example). For galaxies that transition to the kinetic mode before infall, SMBH feedback may have been more important causing their loss of gas.

To summarize, jellyfish galaxies as a population are in the thermal mode of SMBH accretion rate and hence feedback. It is possible that ram pressure may actually be responsible for {\it keeping} the jellyfish in the high-accretion mode, by compressing gas toward the SMBH (see Section~\ref{SEC:RP_AGN}): however, we expect that eventually ram-pressure stripping and complete gas removal will take over, even for galaxies that appear jellyfish at the time of inspection. What is also interesting is the fact that, as we show above, many satellites were inspected or even jellyfish in the past, and due to the accumulated environmental effects, perhaps combined with kinetic SMBH feedback, have since \newpage \noindent lost their gaseous tails. The multiple evolutionary tracks shown in Figure~\ref{A:AGN_FB} demonstrate the different ways in which SMBH feedback and environmental processes can simultaneously affect satellite galaxies. As a corollary, this analysis also reveals that, while it is expected that the fraction of jellyfish galaxies decreases with satellite stellar mass or satellite-to-host mass ratio \citep{yun_jellyfish_2019, zinger_jellyfish_2023}--due to the stronger restoring force against ram pressure of more massive galaxies, the scarcity of jellyfish galaxies in absolute number may actually be enhanced compared to what is expected because of environmental processes only because of their very AGN feedback. At least in TNG, more massive galaxies are more frequently in kinetic mode of feedback and hence have an additional physical avenue for their gas to be depleted, reducing their chances of exhibiting ram-pressure stripped tails than in the absence of feedback.